\documentclass[a4paper,fleqn]{cas-dc}

\usepackage[numbers]{natbib}
\usepackage{amsmath,amssymb,amsfonts}
\usepackage{algorithmic}
\usepackage[ruled,vlined]{algorithm2e}
\usepackage{graphicx}
\usepackage{textcomp}
\usepackage{xcolor}
\usepackage[english]{babel}
\usepackage{setspace}
\usepackage{booktabs}
\usepackage{subcaption}
\usepackage{float}
\usepackage{url}
\usepackage{pgfplots}
\usepackage{pifont}
\pgfplotsset{compat=newest}
\usepackage{enumitem}

\renewcommand{\footnoterule}{%
  \kern -3pt
  \hrule width 0.487\textwidth height 0.4pt
  \kern 2.6pt
}

\usepackage{tikz}
\newcommand*\filledcircled[1]{%
  \tikz[baseline=(char.base)]{%
    \node[
      circle,
      draw=black,        % circle outline
      fill=black,        % circle fill
      text=white,        % digit color
      font=\scriptsize,  % use a smaller font
      inner sep=0.5pt,   % very little padding
      minimum size=1em   % shrink the overall diameter
    ] (char) {#1};
  }%
}
\newcommand*\outlinedcircled[1]{%
  \tikz[baseline=(char.base)]{%
    \node[
      circle,
      draw=black,         % circle outline
      text=black,         % digit color
      font=\scriptsize,
      inner sep=0.5pt,
      minimum size=0.8em
    ] (char) {#1};
  }%
}

\begin{document}
\let\WriteBookmarks\relax
\def\floatpagepagefraction{0.01}
% minimum proportion of the body text on a page
\def\textpagefraction{0.01}

\shorttitle{Locality-Aware Scheduling for Stream Processing Applications in Serverless Environments}
\shortauthors{T. Qi et~al.}

\title [mode = title]{LcSched: Locality-Aware Distributed Scheduling for Stream Processing in Stateful Serverless Environments}

\author{Tianyu Qi}[orcid=0009-0006-1616-9902]
\cormark[1]
\ead{tiqi@student.unimelb.edu.au}

\author{Maria A. Rodriguez}[orcid=0000-0002-2831-8526]
\ead{marodriguez@unimelb.edu.au}

\author{Rajkumar Buyya}[orcid=0000-0001-9754-6496]
\ead{rbuyya@unimelb.edu.au}

\affiliation{organization={The Quantum Cloud Computing and Distributed Systems (qCLOUDS)                                  Lab, School of Computing and Information Systems, The University                               of Melbourne},
                country={Australia}}

\cortext[cor1]{Corresponding author}

\begin{abstract}
Stateful Serverless (Function-as-a-Service) environment are increasingly used for stream processing applications. These applications are often structured as pipelines where data flows through a sequence of distinct processing operators. Many operators are stateful and require access to the processing state during execution. To address this, stateful serverless platforms introduce state servers on worker nodes, co-locating state with computation to dramatically reduce access latency. In serverless, these data pipelines are realized by chaining function invocations, passing the processed data from one function to the next. To reduce network overhead, state-of-the-art scheduling approaches employ a distributed worker-level scheduling model, routing locally generated chained requests directly to destination workers, bypassing centralized schedulers. However, these approaches always co-locate state and execution for the same operator in the same worker, which forces most invocations to the next operator to be transferred across workers, thereby increasing both latency and I/O overhead. As a result, this paper proposes LcSched, a novel distributed scheduling strategy tailored for stream processing applications in stateful serverless environments. LcSched seeks to maximise local execution of chained function invocations while ensuring that each execution accesses its state from the local worker. It also monitors workload at runtime to maintain workload balance between workers. Experimental results show that LcSched improves throughput by up to 3.6x and reduces end-to-end latency by up to 79\% compared to the state-of-the-art distributed scheduling strategy for stateful serverless environments.

\end{abstract}


\maketitle

\section{Introduction}
\label{sec:introduction}

In recent years, serverless computing has gained significant traction as a cloud computing model. It offloads infrastructure management to the cloud provider and supports the rapid scaling of lightweight execution units to accommodate fluctuations in workload volume~\cite{cardellini22streamsurvey}. These properties align closely with the characteristics of big data stream processing applications, which require low-latency execution, elastic scalability, and simplified operational management. Consequently, extensive research has emerged on the use of serverless platforms for streaming applications, including use cases such as video processing~\cite{ao2018serverlessvideo}, data analytics~\cite{nastic2017serverlessstream}, and IoT sensor monitoring~\cite{hall2019execution, cicconetti2020decentralized, yao2023performance}.

Stream processing applications are commonly modeled as Directed Acyclic Graphs (DAGs), where nodes represent data processing operators and edges capture the flow of data~\cite{hirzel2014streamoptimizations}. On serverless platforms, these operators are implemented as functions, with data passed between them through chains of function invocations. A challenge arises when operators are stateful, as their logic depends on previously computed intermediate results~\cite{xu2023statefulapplication}. Because conventional serverless functions are stateless, such state must be fetched from remote storage on every invocation, introducing substantial latency~\cite{yang2023faasctdo}. To improve \textbf{state locality}, stateful serverless platforms retain operator state on worker nodes, enabling local access and thereby reducing latency~\cite{shillaker2020faasm,mvondo2021ofc, romero2021faast}.

\begin{figure}[]
\centering
\includegraphics[width=0.48\textwidth]{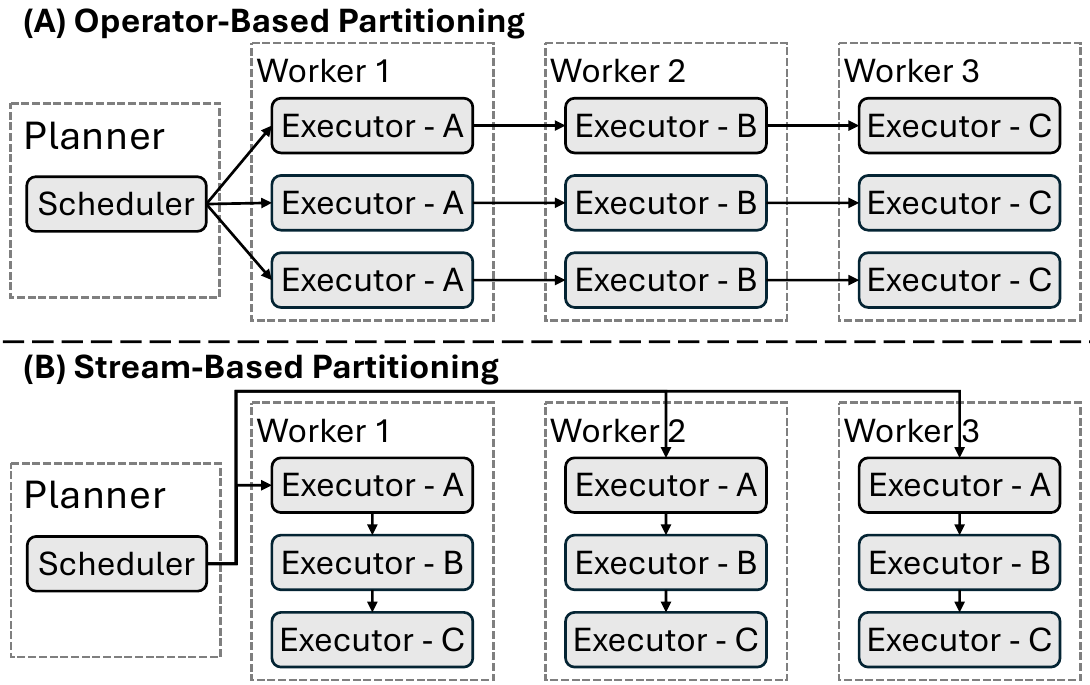}
\caption{Comparison of chained request routing patterns resulting from different application partitioning methods. The application follows a fixed data flow from A to B to C.}
\label{fig:1.schedulemode_compare}
\end{figure}

In serverless environments, DAG applications produce chained requests as operators invoke downstream operators on worker nodes. To reduce end-to-end latency, many platforms adopt a distributed scheduling model, where each worker hosts a scheduling engine that routes chained requests directly to destination workers~\cite{wang2025octopus,zhang2025decentralizedschedulingweb,chen2025ekko}. For stateful serverless platforms, existing distributed scheduling strategies often rely on an \textbf{operator-based} partitioning approach, in which each operator is statically assigned to a specific worker prior to execution~\cite{sreekanti2020cloudburst,li2022faasflow}. This pre-assignment enables scheduling engines to route chained requests directly to their designated workers at runtime. However, as illustrated in Fig.~\ref{fig:1.schedulemode_compare}(A), placing all instances of an operator on a single worker causes most chained requests to be routed across worker boundaries, thereby increasing cross-worker transmission costs for requests\cite{carver2020wukong}. To address this overhead, \textbf{request locality} should be taken into consideration by striving for chained requests to be executed on the same worker where they originate.

To improve request locality, some distributed scheduling strategies adopt a \textbf{stream-based} partitioning approach, where each chained request is executed on the same worker where it originates~\cite{akkus2018sand,kotni2021faastlane,wang2024briskchain}, as shown in Fig.~\ref{fig:1.schedulemode_compare}(B). While the goal of this approach is to reduce cross-worker communication, it has two main disadvantages for stateful applications. First, state locality may be compromised, since locally executed requests may still require access to state stored on remote workers. Second, the number of chained requests generated by each operator can vary significantly with the input data, creating uneven and unpredictable downstream workloads~\cite{nastic2017serverlessstream}. Executing all chained requests locally can therefore lead to severe imbalances across workers. Consequently, the key challenge in stateful serverless platforms for streaming applications is to maximize request locality while simultaneously preserving state locality and maintaining \textbf{workload balance}.

Another key limitation is that current worker-level scheduling engines schedule requests without considering the state that will be accessed by subsequent requests. Incorporating this consideration could improve request locality while preserving state locality. For example, consider a \emph{network monitoring} application that detects whether a successful login is preceded by multiple failed attempts from its host~\cite{gedik2014elastic}. A stateless \emph{parse logins} operator parses and classifies incoming login records and forwards the processed records to a stateful \emph{join} operator, which maintains a per-host login history and detects fraudulent login records based on the history from their host. In stateful serverless platforms, these login histories might be distributed across workers, with each worker managing the state for a disjoint set of hosts. For a login record from host $h_i$, executing its \textit{parse logins} procedure on the worker that holds the login history for $h_i$ ensures that the subsequent chained request to \textit{join} can be executed locally and access its state without cross-worker communication.

To address these challenges, we propose LcSched, a distributed scheduling strategy for stream processing in stateful serverless systems. LcSched makes scheduling decisions that simultaneously account for state locality, request locality, and workload balance to achieve low latency and high throughput. We propose a hybrid partitioning strategy that combines operator-based and stream-based approaches. Our method partitions the application into operator groups based on their connectivity and internal state. This co-location strategy ensures that chained requests within a group are executed on a single worker that holds all the state required for their processing. Each group is then assigned to a dedicated pool of workers, allowing the scheduling engine on each worker to route incoming requests to workers based on a trade-off between request locality and workload balance. At runtime, LcSched continuously monitors worker workloads and dynamically adapts its request routing strategy to maintain workload balance. Across six benchmark applications on clusters of up to ten workers, LcSched improves throughput by up to 3.6x and reduces end-to-end latency by up to 79\% relative to the evaluated FaaSFlow~\cite{li2022faasflow} scheduling strategy.

Our approach makes the following contributions:
\begin{itemize}
    \item \textbf{Downstream state/locality-aware grouping and routing:} We identify operator paths with compatible partitioning semantics and use the state required by downstream invocations to determine where earlier invocations should execute, increasing request locality while preserving state locality.
    
    \item \textbf{Adaptive Runtime reconfiguration:} We monitor per-operator request distributions and adjust routing probabilities to alleviate worker imbalance, selectively sacrificing request locality while retaining local state access.
\end{itemize}

The rest of the paper is organised as follows. Section 2 discusses background concepts, Section 3 presents the design of the proposed LcSched scheduling system, Section 4 evaluates our approach and compares it with the baseline, and Section 5 identifies unique aspects of our work compared to related works. Finally, Section 6 concludes the paper along with directions for future work.

\section{Background}
\label{sec:background}
\begin{figure*}[t]
  \centering
  %── First figure ──
  \begin{minipage}[b]{0.3\textwidth}
    \centering
    \includegraphics[width=\textwidth]{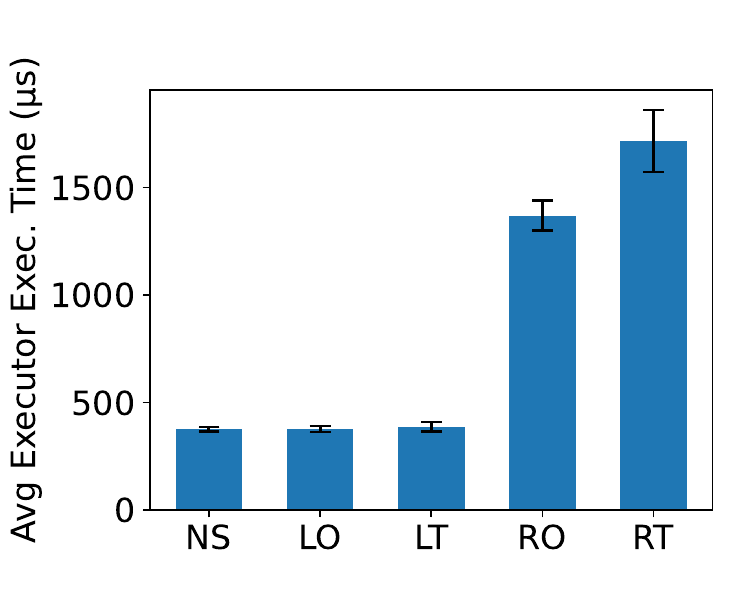}
    \subcaption{Comparison of executor execution latency under different state access circumstances.}
    \label{fig:2.1.state_latency}
  \end{minipage}
  \hfill
  %── Second figure ──
  \begin{minipage}[b]{0.3\textwidth}
    \centering
    \includegraphics[width=\textwidth]{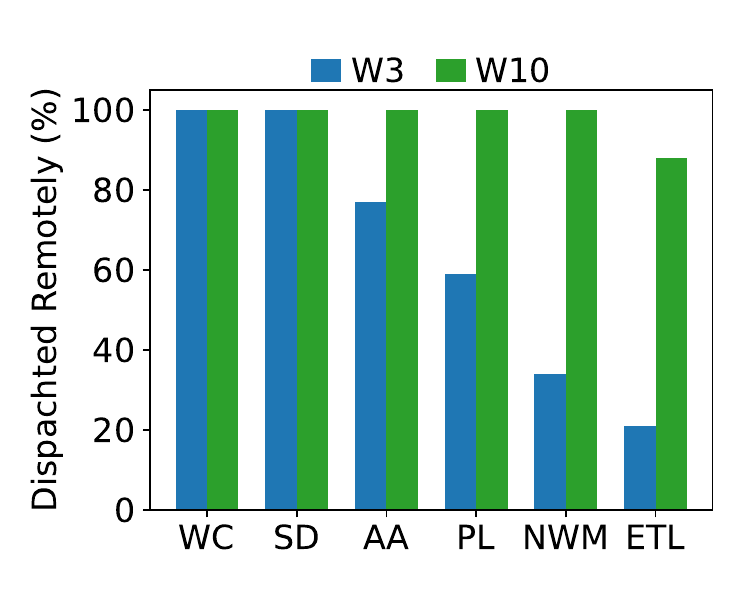}
    \subcaption{Proportion of cross-worker chained requests in FaaSFlow under 3-worker (W3) and 10-worker (W10) clusters.}
    \label{fig:2.2.faasflow_internode}
  \end{minipage}
  \hfill
  %── Third figure ──
  \begin{minipage}[b]{0.3\textwidth}
    \centering
    \includegraphics[width=\textwidth]{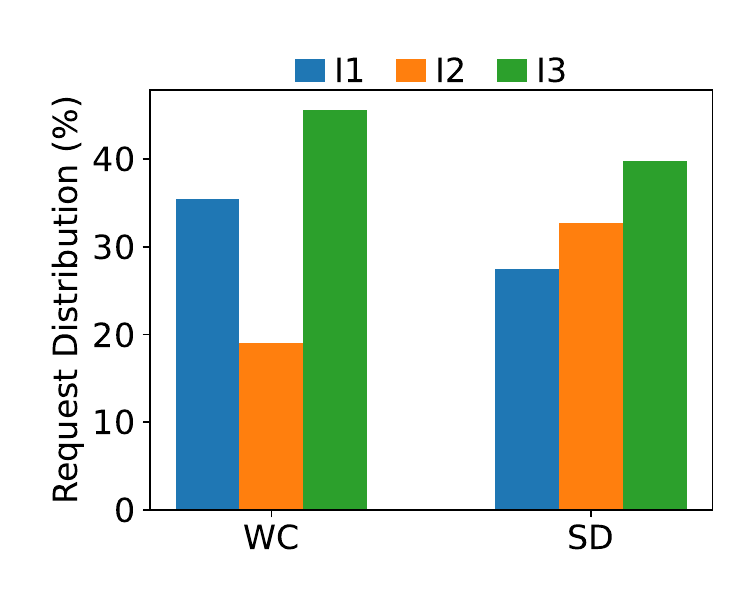}
    \subcaption{Request distribution across three state instances (I1, I2, and I3) for the partitioned stateful operator.}
    \label{fig:2.3.partitioned_key_distribution}
  \end{minipage}
\end{figure*}
\subsection{Operator Types in Stream Processing Applications}
\label{subsec:2.1.streaming_application_type}

In stream processing applications, operators can be classified as stateless, stateful, or partitioned stateful based on their state characteristics~\cite{gedik2014elastic}. A stateless operator processes each request independently, without relying on prior computations. In contrast, a stateful operator maintains intermediate results (i.e., state) that are updated when a request is processed~\cite{carbone17stateflink}. A partitioned stateful operator is a specialized type of stateful operator in which the state is partitioned based on a predefined partitioning attribute~\cite{fernandez2013scalestream}. Each incoming request includes a key for this attribute, and the operator only accesses the corresponding state partition to process the request. For instance, in a word-count application, the \emph{count} operator uses the \emph{word} itself as the partitioning attribute and maintains a separate counter for each distinct word. An invocation request to \emph{count} is represented as \texttt{\{"word": "hello"\}}.

The stateful and partitioned stateful operators maintain their intermediate results as state variables. In serverless platforms, these variables are serialized into an object for persistence on a storage server. In this paper, we term this persisted object as a state instance. For a partitioned stateful operator, to handle higher input rates, a keyed stateful operator can be scaled out by creating multiple state instances, each processing a subset of the input stream. Its state is organized as a key-value map, where the key represents the partitioning attribute and the value holds the corresponding state partition. Increasing the parallelism distributes the key-value pairs of this map across multiple state instances. The parallelism of such an operator can be adjusted as needed. With multiple state instances, all requests sharing the same key are consistently routed to the same instance. LcSched ensures this by implementing a hash-based partitioning strategy that deterministically maps each partitioning attribute key to a specific instance.

The parallelism for a stateful operator is typically predefined by users and fixed, as it is often constrained by the application's logic. For instance, an operator that tracks the most frequent words across an entire text stream must maintain a single global state instance; increasing its parallelism would partition the state and produce incorrect results. 

% As an example, the \emph{total count} operator in the \emph{log processing} application records the total number of logs processed~\cite{stromax}. With one parallelism, its state is a single counter. If its parallelism were increased, this counter would be replicated, and each replica would only count a subset of the input stream. This approach violates the operator's fundamental design, and therefore, its parallelism cannot be increased.

\subsection{State Locality}
\label{subsec:2.2.state_locality}

Stateful and partitioned stateful operators read their state before processing each request and write back the updated state upon completion, making execution latency highly sensitive to state access time. This motivates the notion of state locality: an invocation has state locality when it executes on the worker that stores the state instance it accesses. To quantify this effect, we designed five microbenchmark operators and ran them under the experimental environment in (\S \ref{sec:evaluation}). All microbenchmarks implement identical no-op function logic but differ only in their state‐access pattern, as summarized in Tbl.~\ref{tab:2.operator_state_access}.

\begin{table}[ht]
  \centering
  \caption{Microbenchmarks and their state‐access patterns}
  \label{tab:2.operator_state_access}
  \begin{tabular}{ll}
    \hline
    MB & State Access Pattern \\ 
    \hline
    NS & no state access\\
    LO & read state from local state server\\
    LT & read \& write on local state server\\
    RO & read state from state server in another worker\\
    RT & read \& write on state server in another worker\\
    \hline
    \multicolumn{2}{l}{\footnotesize MB: Microbenchmarks} \\
  \end{tabular}
\end{table}

By processing only one request at any given time, we compared their execution latency—from the time the operator's executor starts until completion. Fig.~\ref{fig:2.1.state_latency} shows that the NS, LO, and LT benchmarks each complete in around 400 $\mu$s, indicating that the latency introduced by local state access is negligible. By contrast, when accessing state remotely, the RO benchmark incurs approximately 1,200 $\mu$s of latency, and the RT benchmark further increases this to around 1,400 $\mu$s—about a threefold increase over local state access. These benchmarks use a one-byte value as state. However, prior work has shown that, as state size increases, the latency overhead of remote state accesses relative to local accesses becomes increasingly pronounced~\cite{sreekanti2020cloudburst}.

\subsection{Request Locality}
\label{subsec:2.3.request_locality}

In distributed scheduling models, transferring requests between workers incurs extra latency and communication overhead~\cite{carver2020wukong}. This motivates the notion of request locality: an invocation has request locality when it is executed on the same worker that generated it, avoiding cross-worker request transmission. In our experiment environment (\S~\ref{sec:evaluation}), a local request transfer takes approximately 0.1 ms, whereas a cross-worker transfer takes around 1 ms. However, request locality is undermined in operator-based partitioning approaches, which co-locate all state and execution logic of an operator on the same worker. This leads to different operators being executed on separate workers, thereby forcing chained requests transmitted across workers. FaaSFlow~\cite{li2022faasflow} provides a popular distributed scheduling strategy based on operator-based partitioning for stateful serverless platforms. We evaluate FaaSFlow by running our benchmarks on both 3-worker and 10-worker clusters. The benchmarks comprise six stream processing applications, ranging from 2 to 10 operators: Word Count (WC), Spike Detection (SD), Advertisement Analytics (AA), Page Load (PL), Network Monitor (NWM), and Extraction, Transform and Load (ETL). Full testbed and workload details are provided in \S\ref{sec:evaluation}. 

As shown in Fig.~\ref{fig:2.2.faasflow_internode}, where applications are ordered left to right by their number of operators, a clear trend emerges on the 3-worker cluster. Applications with more operators exhibit a lower proportion of cross-worker request routing. This is because, when the number of operators exceeds the number of workers, multiple adjacent operators are assigned to the same worker, so chained requests between them execute locally. When the cluster size increases to 10 workers, the majority of requests are routed across workers for execution. In the evaluated FaaSFlow placement, cross-worker routing is common for small pipelines and becomes more frequent as the number of workers increases.

\subsection{Workload Variability}

Stream-based approaches execute all chained requests on the worker where they originate~\cite{kotni2021faastlane, yu2023pheromone}. However, this can lead to severe imbalances across workers for two reasons.

\noindent\subsubsection{Bursty Downstream Workloads} Downstream workloads are often unpredictable and bursty~\cite{nastic2017serverlessstream}. A common example for this is an aggregation operator, which buffers requests and then flushes them to a successor in a burst once a threshold (e.g., a time window or item count) is met~\cite{bordin2020dspbench}. This behavior creates a large batch of chained requests when the threshold is triggered, while no requests are produced at other times. 

\noindent\subsubsection{Request Skew} While each request accesses a specific state instance, the overall distribution of these requests is often highly skewed. To demonstrate this, for Spike Detection (SD) and Word Count (WC) applications from \S\ref{sec:evaluation}, we configured their partitioned stateful operators with three state instances. We then processed an input stream from a real-world trace and measured the distribution of requests across the different state instances~\cite{bordin2020dspbench}. Fig.~\ref{fig:2.3.partitioned_key_distribution} illustrates a markedly uneven distribution of requests across state instances. The skew results from the non-uniform distribution of real-world input data, which causes some state instances to be accessed far more frequently than others. When requests are executed on the worker that contains the state instance they access, this skewed request distribution directly translates into an imbalanced workload across workers. While this issue could be addressed through dynamic state migration, this approach introduces substantial runtime complexity and is beyond the scope of this work. However, when scheduling their chained requests, the scheduler must therefore increase request locality without allowing key skew or downstream burstiness to overload individual workers.

\section{LcSched Scheduling System Design}
\label{sec:design}

\begin{figure}[]
\centering
\includegraphics[width=0.48\textwidth]{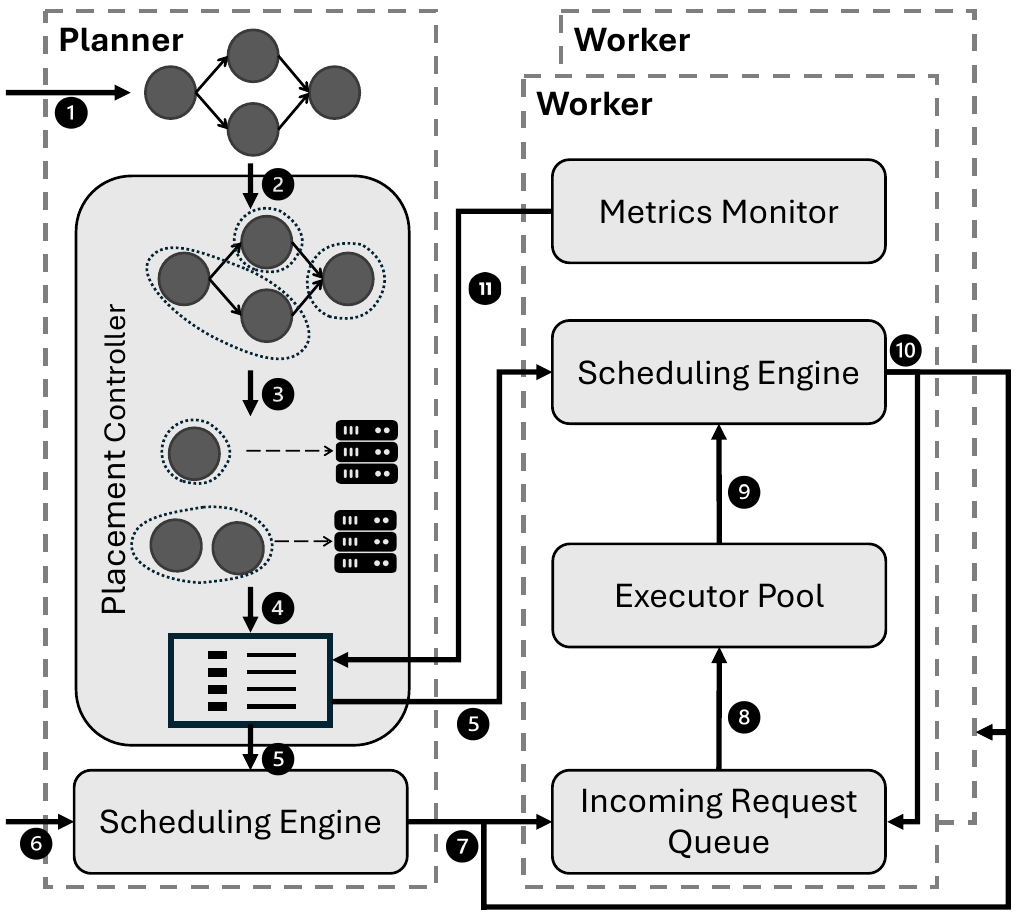}
\caption{Architecture of LcSched}
\label{fig:3.method_structure}
\end{figure}

Fig.~\ref{fig:3.method_structure} illustrates the architecture of our design. The planner hosts a placement controller that maps application operators to workers and a scheduling engine that dispatches input requests. Each worker provides an incoming request queue, an executor pool, a scheduling engine for locally generated chained requests, and a runtime metrics monitor. Each planner and worker runs on a dedicated host node.

\subsection{High-Level Design}

The workflow of LcSched is illustrated in Fig.~\ref{fig:3.method_structure}. Before execution, the user registers the application with the placement controller~\filledcircled{1}. The controller then partitions the application into operator groups~\filledcircled{2} and assigns these groups to workers~\filledcircled{3}. For any group assigned to multiple workers, the controller distributes its execution share among workers~\filledcircled{4}. Next, the placement controller propagates the result to the scheduling engines~\filledcircled{5}. After initialization, the planner’s scheduling engine receives input requests~\filledcircled{6} and dispatches them to workers’ incoming request queues~\filledcircled{7}. Inside each worker, after executing requests from the queue~\filledcircled{8}, the executor sends generated chained requests to the local scheduling engine~\filledcircled{9}. Each scheduling engine then routes these chained requests to the destination workers~\filledcircled{10}. During runtime, the metrics monitor tracks each operator’s workload, such as the number of processed requests, and periodically reports metrics back to the placement controller~\filledcircled{11}, enabling dynamic reconfiguration of execution shares.

\subsection{Placement Controller Partitioning and Mapping}

Firstly, the user registers all operators~$O$ of application~$A$ with the scheduler. For each operator~$o$, the user provides its name~$o.\mathrm{name}$, type~$o.\mathrm{type}$ (i.e., Stateless, Stateful, or Partitioned Stateful), input attributes~$o.\mathrm{attr}$  (e.g., in a Word Count application, the input attribute of the count operator is $\texttt{"word"}$, meaning each input request carries a \texttt{"word"} key-value pair, such as $\{\texttt{"word":"dog"}\}$), and successor operators~$o.\mathrm{succ}$. Partitioned-stateful operators additionally specify a partitioning attribute $o.\mathrm{prt}$, whereas stateful operators additionally specify the desired parallelism $o.\mathrm{para}$. 

\begin{figure}[]
\centering
\includegraphics[width=0.48\textwidth]{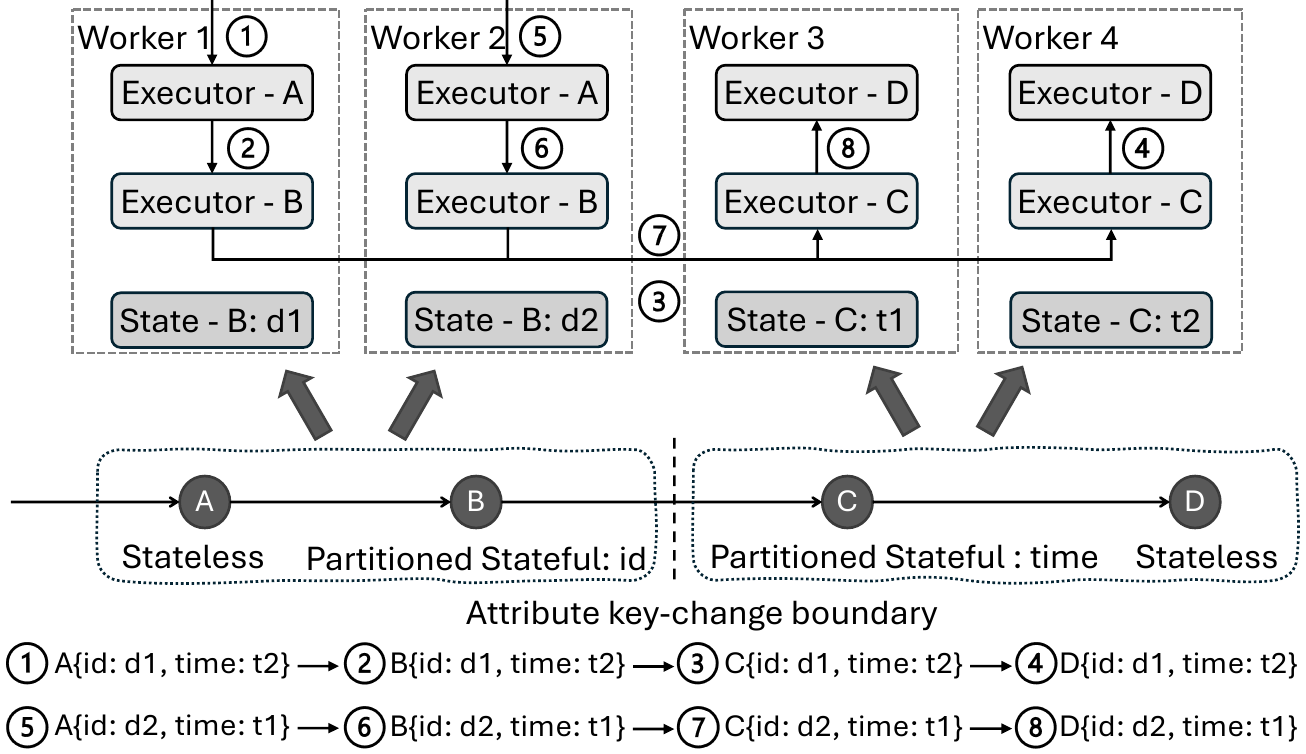}
\caption{An example application consisting of four operators: A (Stateless), B (Partitioned Stateful), C (Partitioned Stateful), and D (Stateless). The partitioning attributes for B and C are id and time, respectively. The operators are organized into two groups, with A and B assigned to Workers 1 and 2, and C and D assigned to Workers 3 and 4. The state distribution is shown at the bottom of each worker. For example, \{State – C: t1\} represents the state of operator C corresponding to time t1. This figure illustrates the execution paths of two distinct input requests~\protect\outlinedcircled{1} and~\protect\outlinedcircled{5}, and their subsequent chained requests within our proposed method.}
\label{fig:3.1.schedule_pattern}
\end{figure}

After receiving the application's information, the scheduler partitions it into operator groups. The routing of requests between operators within the same group then follows three key principles, as illustrated in Fig.~\ref{fig:3.1.schedule_pattern}:
\begin{itemize}[leftmargin=*]
    \item \textbf{Group Co-location:} The requests to the operators in the same group are executed by a dedicated subset of workers. For example, requests to operators A and B are only executed on workers 1 and 2.

    \item \textbf{State Locality:} Requests are executed on the worker that holds the state they act upon. In the figure, all requests to operators B and C access their state locally.

    \item \textbf{Intra-Group Request Locality:} Chained requests between operators within the same group are always executed locally.  For instance, all chained requests between A and B, and between C and D, are executed on the worker where they originate. To preserve state locality at the same time, LcSched groups partitioned stateful operators only along paths that preserve the partitioning attribute required by the downstream partitioned-stateful operator. An operator that transforms, removes, or replaces this key creates a group boundary. For example, because B and C partition on different attributes, they are placed in separate groups.
    
\end{itemize}

In our design, when a partitioned stateful operator is assigned to multiple workers, one state instance is placed on each worker. Additionally, for different partitioned stateful operators within the same group, if they share the same partitioning attribute, the state partitions corresponding to the same key reside on the same worker.

\RestyleAlgo{ruled}
\begin{algorithm}
\caption{ApplicationPartition}
\label{alg:application_partition}
\LinesNumbered

% \KwData{}

\SetKwInput{KwTypes}{O.type}
\KwTypes{\texttt{SL} (stateless), \texttt{PSF} (partitioned stateful)}

\DontPrintSemicolon
\SetKwProg{Fn}{Function}{:}{}
\SetKwFunction{Eligible}{Eligible}
\SetKwFunction{Group}{Group}

\Fn{\Eligible{$grp, o$}}{
  \lIf{$o.type=\texttt{SL}$}{\KwRet true}
  \uIf{$o.type=\texttt{PSF}$}{
    \ForEach{$m\in grp$}{
    \uIf{$m.t=\texttt{SL}$ \textbf{and} $o.prt \notin m.{attr}$}{\KwRet false}
    \uIf{$m.t=\texttt{PSF}$ \textbf{and} $o.prt\ne m.prt$}{\KwRet false}
    }
    \KwRet true
  }
  \KwRet false
}
\Fn{\Group{$u$}}{
  $stack\gets [u]$, $grp\gets[\,]$\;
  \While{$stack\neq\emptyset$}{
    $o\gets stack.pop()$\;
    \lIf{$o\in vis$}{continue}
    $vis.add(o)$, $grp.push(o)$\;
    $S \gets$ successors of $o$ sorted by chained-request rate (asc)\;
    \ForEach{$o_{succ}\in S$}{
      \uIf{\Eligible{$grp,o_{succ}$}}{$stack.push(o_{succ})$}
      \uElseIf{$o_{succ}\notin vis$}{\Group{$o_{succ}$}}
    }
  }
  \lIf{$grp\neq\emptyset$}{$G.push(grp), u.{head} \leftarrow true$}
}
\BlankLine
$G\gets[\,]$, $vis\gets\emptyset$\;
\ForEach{input operator $u$ in $A$}{\Group{$u$}}
\KwRet G
\end{algorithm}

The scheduler partitions the application graph into operator groups using a Depth-First Search (DFS) algorithm, as shown in Alg.~\ref{alg:application_partition}. Since in our design, intra-group chained requests are always executed locally, to maximize request locality, we construct maximal groups subject to state-compatibility and key-preservation constraints, while allowing group boundaries where larger groups would violate correctness or reduce placement feasibility. The process begins by initiating a DFS from each input operator (i.e., one that receives input from the source directly) (Lines~24--26). The algorithm expands a group $grp$ by adding eligible (i.e., \textsc{Eligible} returns true) successor operators and then continuing the expansion from those newly added successors (Lines~11--23). Because an operator can have multiple successors, to ensure that all downstream branches are explored, the scheduler pushes all eligible successor operators into a stack $stack$ (Lines~19--20), and iterates until the stack is empty. For each eligible operator in the stack, if it has not been visited, this operator is added to the current group (Lines~15--16). If added, the scheduler then examines its successors. It first sorts the successors in ascending order of their chained-request rate from the current operator, so that, due to the stack's last-in-first-out order, the most heavily-communicating successor is explored first (Line~17). Then, for each successor, if \textsc{Eligible} holds, it is pushed onto the stack (Lines~19--20); otherwise, a new operator group is initiated for that successor (Lines~21--22). Once the stack becomes empty, the current group is appended to the group set, and the start operator of the current group is marked as the head operator (Line~23).

\begin{figure}[]
\centering
\includegraphics[width=0.48\textwidth]{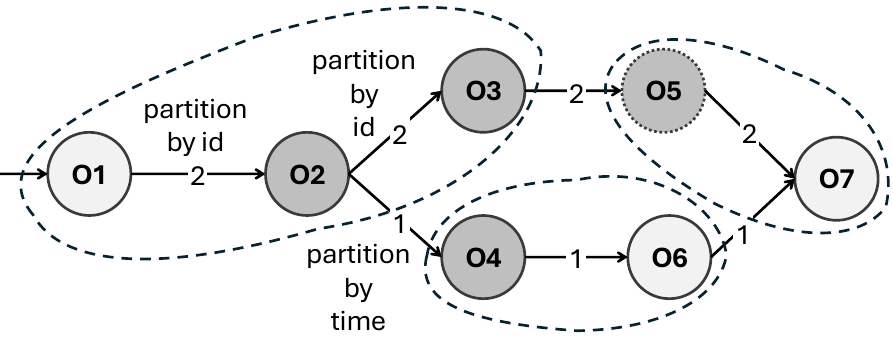}
\caption{Illustrative example of graph partitioning. Stateless operators are shown in light grey, partitioned stateful operators in dark grey, and (non-partitioned) stateful operators in dark grey with a dotted outline. The partitioning yields three groups: $\{O1, O2, O3\}, \{O5, O7\}, \{O4, O6\}$. O4 starts a new group because it is partitioned by a different attribute, and O5 starts a new group because it is stateful.}
\label{fig:2.1.partition_example}
\end{figure}

The decision of whether a successor is eligible is based on its type (Lines~1--10): If the successor operator is stateless, it is directly added to the current group (Line~2). If the successor operator is partitioned stateful, it will be added only if two conditions are satisfied: (i) all stateless operators within the current group include the successor’s partitioning attribute in their input attributes (Lines~5--6), and (ii) all partitioned stateful operators in the current group use the same partitioning attribute as the successor (Lines~7--8). If the successor is stateful, it is not eligible (Line~10). An example of the partitioning is illustrated in Fig.~\ref{fig:2.1.partition_example}.

The operator $o$ that initiates a new group is designated as the \textbf{head operator} $o_{head}$ in this work. For any stateless operator $o_{sl}$ that is upstream of a partitioned stateful operator $o_{psf}$ within the same group, we define it as \textbf{having affinity with} $o_{psf}$. This means that when scheduling a request to $o_{sl}$, the scheduler should consider the location of the state that the subsequent chained request to $o_{psf}$ will access, so that both requests can be executed on the same worker. We refer to the worker holding this state partition as the \textbf{state-aligned worker} $w_s$ for that request. It is worth noting that when a stateless operator has multiple partitioned stateful operators downstream in the same group, these operators necessarily share the same partitioning attribute, as this is a precondition for grouping them together. In this case, the scheduler uses the first partitioned stateful operator to determine the affinity.

Next, the scheduler maps groups across all available workers according to worker capacities and operator demands. Let $w.\mathrm{cap}$ denote the capacity (e.g., CPU cores) of worker $w$. To determine how much of the cluster's capacity each operator requires, we define the required capacity for each operator $o$ as follows. For operator $o$, let $o.\mathrm{proc}$ be the number of processed requests, $o.\mathrm{creq}$ be the number of chained requests it generates, and $o.\mathrm{exec}$ be the average execution time of operator $o$ (all measured over a time window). We define operator $o$'s workload $o.\mathrm{load}$, i.e., the total processing time it consumes, as: 
\[o.\mathrm{load}=o.\mathrm{proc} \times o.\mathrm{exec}\]
Let $\alpha>0$ be a weighting factor, given an application with $M$ operators and a cluster with $N$ workers, the required capacity for operator $o$ is:
\[
o.cap = \frac{o.load + \alpha\times o.creq}{\sum_{j=1}^{M} {o_j.load} + \alpha \times \sum_{j=1}^{M} o_j.creq} \times \sum_{i=1}^{N} w_i.cap
\]
It is worth noting that an operator’s parallelism does not affect capacity allocation. Workers spawn concurrent executors to process requests belonging to each parallelism instance, and the maximum executors per instance is determined by the operator’s required capacity divided by its parallelism degree.

The scheduler assigns operator groups to workers using a bin-packing strategy. It computes an expected share for each group's required capacity on each worker — representing the fraction of the group's workload that worker is expected to handle. The required capacity of each group is calculated by summing the individual capacities of its constituent operators. It sequentially places each group into workers: if a worker has enough remaining capacity, the whole group is placed there; otherwise, it fills the worker’s remaining capacity and continues on to the next worker until the group’s demand is met. This procedure produces the expected capacity share of each group $g$ across workers $W$, denoted as $S^\mathrm{exp}_{g}(W)$. Here, $S^\mathrm{exp}_{g}(w)$ represents the fraction of $g$’s required capacity that is assigned to worker $w$. Since all operators in a group share the same capacity distribution, each operator $o \in g$ inherits this pattern, i.e., $S^\mathrm{exp}_{o}(W) = S^\mathrm{exp}_{g}(W)$.

Then, the scheduler assigns each operator's state instances to workers, thereby allowing the scheduling engines to route requests directly to the state they act upon. For each stateful operator $o_{sf}$, the user-specified parallelism remains fixed. The scheduler distributes the operator's state instances among workers, with the allocation for a worker $w$ proportional to its expected share $S^\mathrm{exp}_{o_{sf}}(w)$. For each partitioned stateful operator $o_{psf}$, its parallelism is adjusted to the number of workers this operator is assigned to, with each worker holding exactly one state instance. When mapping partitioning attribute keys to state instances, we use consistent hashing with virtual nodes. The number of virtual nodes placed on the instance in $w_i$ is proportional to $S^\mathrm{exp}_{o_{psf}}(w_i)$. This ensures that key distribution across workers aligns with the planned capacity allocation. Finally, the scheduler propagates the expected shares and state placements to scheduling engines.

\subsection{Scheduling Engine}

For requests to stateful or partitioned stateful operators, the scheduling engine routes them to the worker that contains the state instance they act upon. For requests to stateful operators, if they do not specify a particular state instance, they are distributed across state instances in a round-robin manner. For requests to partitioned stateful operators, the scheduling engine uses the consistent hashing algorithm on its partitioning attribute to determine the target state instance. For requests to stateless operators, the scheduling engine selects a worker from the assigned set, aiming to maximize request locality while preserving workload balance.

To mitigate the high scheduling overhead on workers, the scheduling engine adopts a probabilistic scheduling model for requests to stateless operators. For a request to an operator $o$ that has no affinity, the scheduling engine dispatches it to worker $w_i$ with probability $P^\mathrm{sch}_o(w_i)$. For a request to an operator $o$ that has affinity with a partitioned stateful operator $o_{psf}$, the scheduling engine inspects the value, $v$, of its input attribute that corresponds to the partitioning attribute of $o_{psf}$. To maximize the request locality, this request is preferably routed to the state-aligned worker $w_s$ that holds the state partition corresponding to $v$. Consequently, the scheduling must account for the state-aligned worker, and the probability of scheduling a request targeting $o$ whose state-aligned worker is $w_s$ to worker $w_i$ is denoted as $P^\mathrm{sch}_{o,w_s}(w_i)$.

For requests to stateless operators, the scheduling policy after application initialization is as follows:
\subsubsection{Scheduling for Head Operator} Requests to head operators originate from an external source or from an operator in another group. For these requests, scheduling decisions disregard request locality and depend on whether the operator has an affinity with another operator:

\noindent \textbf{Head Without Affinity:} Requests are distributed across assigned workers in proportion to their expected shares. Therefore, the scheduling probability of scheduling a request for operator $o$ to a worker $w$ is equal to its expected share,  $P^\mathrm{sch}_o(w) =S^\mathrm{exp}_o(w) \quad \forall w\in W$.

\noindent \textbf{Head With Affinity:} The request is deterministically routed to its state-aligned worker. For example, in Fig.\ref{fig:3.1.schedule_pattern}, the operator A is a head operator that has affinity with operator B. Consequently, all requests to operator A are routed to the worker that holds the state partition corresponding to the request's \emph{id} attribute. This means that if the state-aligned worker for a request to operator $o$ is $w_s$, the scheduling probability for that worker is 1 (and 0 for all others): $P^\mathrm{sch}_{o,w_s}(w_s) = 1$.

\subsubsection{Scheduling for Non-Head Operator} Chained requests to a non-head operator originate from other operators in the same group and are always executed locally for request locality. For example, in Fig.~\ref{fig:3.1.schedule_pattern}, the operator D is a non-head operator, and all chained requests to it are executed locally. For any request to operator $o$ generated on worker $w_i$, if $o$ has no affinity, the scheduling probability is $P^\mathrm{sch}_o(w_i) = 1$. If $o$ has affinity with another operator, any request to $o$ is generated on its state-aligned worker $w_s$, as the scheduling engine has already put its preceding request towards the head operator to $w_s$. When routing locally, the request is always scheduled to this state-aligned worker, i.e., $P^\mathrm{sch}_{o,w_s}(w_i) = 1\quad\textbf{where}\quad w_i=w_s$. 

% \noindent\emph{Worker-pool exception.} Suppose operator~$C$ has two predecessor operators~$A$ and $B$. If $C$ is grouped with $A$ and mapped on $w_i$, while $B$ is in another group mapped to $w_j$. For the chained requests to $C$ generated on $w_j$, the scheduling engine on $w_j$ treats them as requests to head operator.

\subsection{Runtime reconfiguration}

To maintain a balanced workload across workers despite unpredictable workload variability, the scheduling engine dynamically adjusts its scheduling probabilities. 

First, to understand the current request distribution, a metrics monitor on each worker logs processed requests per operator. The placement controller then gathers these records to calculate the observed shares $S^\mathrm{obs}_o(w)$ for each operator $o$, which is the fraction of requests to operator $o$ handled by worker $w$ over the last sliding time window. 

Next, the placement controller adjusts request shares accordingly. For each operator $o$, the scheduling engine maintains an adjusted share $S^\mathrm{adj}_o(W)$, initialized with the expected share. This adjusted share is periodically updated using a learning rate $\eta\in(0,1]$ based on the discrepancy between expected and observed request shares: 
\[
S^\mathrm{adj}_{o}(w)= \operatorname{Norm}_{W}\!\Big(S^\mathrm{adj}_o(w)+\eta \big(S^\mathrm{exp}_{o}(w)-S^\mathrm{obs}_{o}(w)\big)\Big)
\]
where $\operatorname{Norm}_{W}$ renormalizes the adjusted shares such that $\sum_{w\in W} S^\mathrm{adj}_o(w)=1$. This update rule increases the adjusted share for any worker whose observed share is below its expected share, and decreases it if above.

Based on the updated adjusted share, the scheduling engine determines the set of underloaded workers $U_o$ for each operator $o$, as:
\begin{equation*}
    U_o = \{w \in \mathcal{W} \mid S^\mathrm{adj}_o(w) > S^\mathrm{exp}_o(w)\}
\end{equation*}
The underloaded worker $w \in U_o$ should receive more requests than currently planned. We define this gap as the deficit $d_o(w)$, which quantifies how much additional workload worker $w$ should accommodate. We define operator $o$'s deficit $d_o(w)$ in worker $w$ and the total deficit $D_o$ for all workers as follows:
\begin{align*}
    d_o(w) & \leftarrow S^\mathrm{adj}_{o}(w) - S^\mathrm{exp}_{o}(w) \\
    D_o & \leftarrow \sum_{w \in U_o} d_o(w)
\end{align*}

Next, the scheduling engine adjusts routing probabilities for each operator based on its type. 
\subsubsection{Reconfiguration for Operator Without Affinity} Scheduling probabilities are updated according to an operator's adjusted and expected shares.

\noindent \textbf{Head Without Affinity:} Requests are always distributed across assigned workers in proportion to their expected shares. In this case, the observed shares consistently match the expected shares, so no adjustments are necessary.

\noindent \textbf{Non-Head Without Affinity:} Requests to these operators are initially executed locally. However, as the downstream requests are generated irregularly, the observed shares can become skewed. For requests to operator $o$ generated on $w_i$, if $w_i$ is in the underloaded set $U_o$, the scheduling engine routes requests exclusively to itself, i.e., $P^\mathrm{sch}_o(w_{i}) = 1$. Otherwise, if $w_i$ is overloaded, the scheduling engine schedules a portion of requests locally and distributes the remainder among underloaded workers in proportion to their deficits:
\[
P^\mathrm{sch}_o(w) =
\begin{cases}
\dfrac{S^\mathrm{adj}_o(w)}{S^\mathrm{exp}_o(w)}, & w = w_{i},\\[6pt]
\bigl(1 - P^\mathrm{sch}_o(w_{i}))\times\dfrac{d_o(w)}{D_o}, & w \neq w_i.
\end{cases}
\]

\subsubsection{Reconfiguration for Operator with Affinity} For requests to an operator $o$ that has affinity with another operator, $P^\mathrm{sch}_o(w)$ denotes the proportion of requests routed to worker $w$. It is determined in the same way as for the operator that has no affinity. Specifically, for a head operator that has affinity, $P^\mathrm{sch}_o(w) = S^\mathrm{exp}_o(w)$; for a non-head operator that has affinity, its $P^\mathrm{sch}_o(w)$ is calculated in the same manner as that of a non-head operator without affinity.

Since each request has a state-aligned worker, the scheduling engine should route it preferentially there. For this purpose, each scheduling engine records $C^\mathrm{aff}_o(w)$—the normalized distribution of locally generated chained requests to $o$ whose state-aligned worker is $w$ (measured over sliding time window). Using this, the scheduler identifies the state-aligned underloaded set $U^\mathrm{aff}_o$ for each operator $o$. This set indicates that, on current worker, the scheduling proportion of operator $o$ to these workers exceeds the proportion of locally generated requests whose state-aligned workers are among them:
\begin{equation*}
    U^\mathrm{aff}_o = \{w \in \mathcal{W} \mid  P^\mathrm{sch}_o(w) > C^\mathrm{aff}_o(w)\}
\end{equation*}
The state-aligned deficit for each worker $w \in U^\mathrm{aff}_o$ denoted as $d^\mathrm{aff}_o(w)$, and the total deficit $D^\mathrm{aff}_o$ are then calculated as:
\begin{align*}
    d^\mathrm{aff}_o(w) & \gets P^\mathrm{sch}_o(w) - C^\mathrm{aff}_o(w) \\
    D^\mathrm{aff}_o & \leftarrow \sum_{w \in U^\mathrm{aff}_o} d^\mathrm{aff}_o(w)
\end{align*}
In this case, if the requests have state-aligned worker $w_s$ and $w_s$ is in state-aligned underloaded set, they are directly routed to $w_s$, $P^\mathrm{sch}_{o,w_s}(w_s) = 1$. Otherwise, only a portion of the requests with state-aligned worker $w_s$ are routed to $w_s$, while the remaining requests are distributed among state-aligned underloaded workers in proportion to their deficits: 
\[
P^\mathrm{sch}_{o,w_s}(w) =
\begin{cases}
\dfrac{P^\mathrm{sch}_o(w)}{C^\mathrm{aff}_o(w)}, & w = w_s,\\[6pt]
\bigl(1 - P^\mathrm{sch}_{o,w_s}(w_s))\times\dfrac{d^\mathrm{aff}_o(w)}{D^\mathrm{aff}_o}, & w \neq w_s.
\end{cases}
\]

For example, for operator $o$ with affinity, when $P^{\mathrm{sch}}_o$ for worker {A, B, C} are {0.6, 0.2, 0.2}, and the recorded $C^\mathrm{aff}_o$ to worker {A, B, C} are {0.8, 0.1, 0.1}. For the request whose state-aligned worker is A, the $P^\mathrm{sch}_{o,w_A}(w)$ to {A, B, C} would be {0.75, 0.125, 0.125}. If request’s state-aligned worker is B (resp. C), it is scheduled exclusively to B (resp. C) with probability 1.

\begin{figure}[]
\centering
\includegraphics[width=0.35\textwidth]{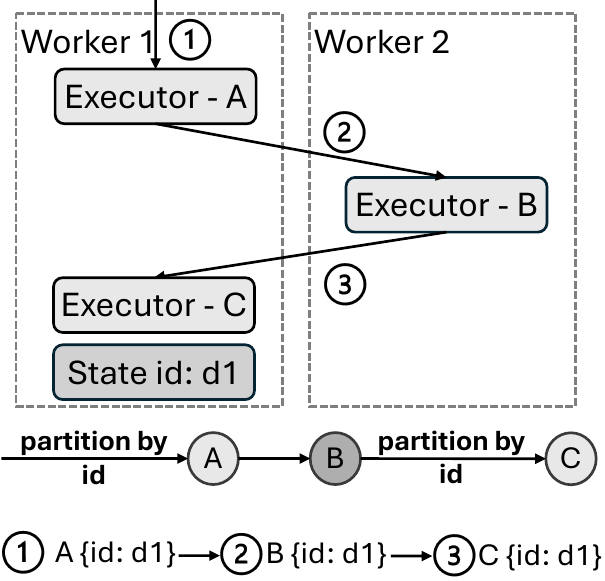}
\caption{An illustrative example where runtime scheduling is exempted. In this three-operator topology, A and C are partitioned stateful operators partitioned by id, while B is stateless. When the state of id \textit{d1} resides on worker 1, B's request is not reconfigured at runtime; otherwise it might incur a double cross-worker request transfer (see figure).}
\label{fig:2.2.runtime_example}
\end{figure}

It is worth noting that runtime reconfiguration does not apply when an operator lies between two partitioned stateful operators partitioned by the same key within the same group. In this case, moving a request of the reconfigurable operator to another worker for load balancing forces its chained request to return to the original worker to access state locally, resulting in double cross-worker request transfers. An example is illustrated in Fig.~\ref{fig:2.2.runtime_example}.

\section{Performance Evaluation}
\label{sec:evaluation}

\noindent\textbf{Implementation.} LcSched is implemented on top of Faasm~\cite{shillaker2020faasm} and extended with two performance-enhancing features: (i) an incoming request queue on each worker to support batch processing, and (ii) a scheduling engine on each worker to enable distributed scheduling. In our system, all chained requests scheduled to the local worker are stored and transferred through shared memory to reduce communication latency. 

\subsection{Experiment Setup}
\subsubsection{Serverless Baselines and Testbed}
All experiments are conducted on a Kubernetes cluster with 11 nodes. One planner node runs on an 8-core AMD EPYC 2.30 GHz machine with 32 GB of RAM, while each of the ten worker nodes runs on a 4-core AMD EPYC 2.30 GHz machine with 16 GB of RAM.

We compare our method against three baselines. The first is the distributed scheduling strategy from FaaSFlow~\cite{li2022faasflow}, which uses an operator-based partitioning approach to map operators to workers. The second baseline employs the same operator mapping approach as FaaSFlow but runs in a centralized scheduling mode. The third baseline is a stream-based approach from Pheromone~\cite{yu2023pheromone}, which preferentially schedules chained requests on the local worker when executors are available. In this baseline, operator state is stored in remote storage, implemented using Redis in this work. For all methods except Pheromone, any request that acts upon a state is always collocated with the worker holding that corresponding state, thereby eliminating the confounding variable of remote state access latency. Since all operator-based methods rely on runtime statistics to evaluate the resources required by each operator and map them effectively to workers, we first run a 10,000-request warm-up phase. Based on the collected statistics, we then remap the operators and begin recording performance. In serverless, a significant portion of CPU time is consumed by loading and reusing executors. Therefore, we set the batch size to 20 in our experiments for all methods, as larger batch sizes can amortize this overhead and increase throughput~\cite{wu2023faasbatch}. Batching affects only the execution layer; the scheduler continues to handle requests individually. It does not affect the fairness of the comparison across different scheduling baselines. Our method is configured with these default parameters: a weighting factor $\alpha=0.1$, a learning rate of $\eta = 0.3$, and a reconfiguration period of 10s. Empirical results indicate that these defaults perform robustly across diverse applications and workloads. FaaSFlow and Pheromone do not employ any parameter configuration.

\subsubsection{Evaluation Metrics and Benchmark Applications}

\begin{figure*}[htbp]
\centering
\includegraphics[width=0.95\textwidth]{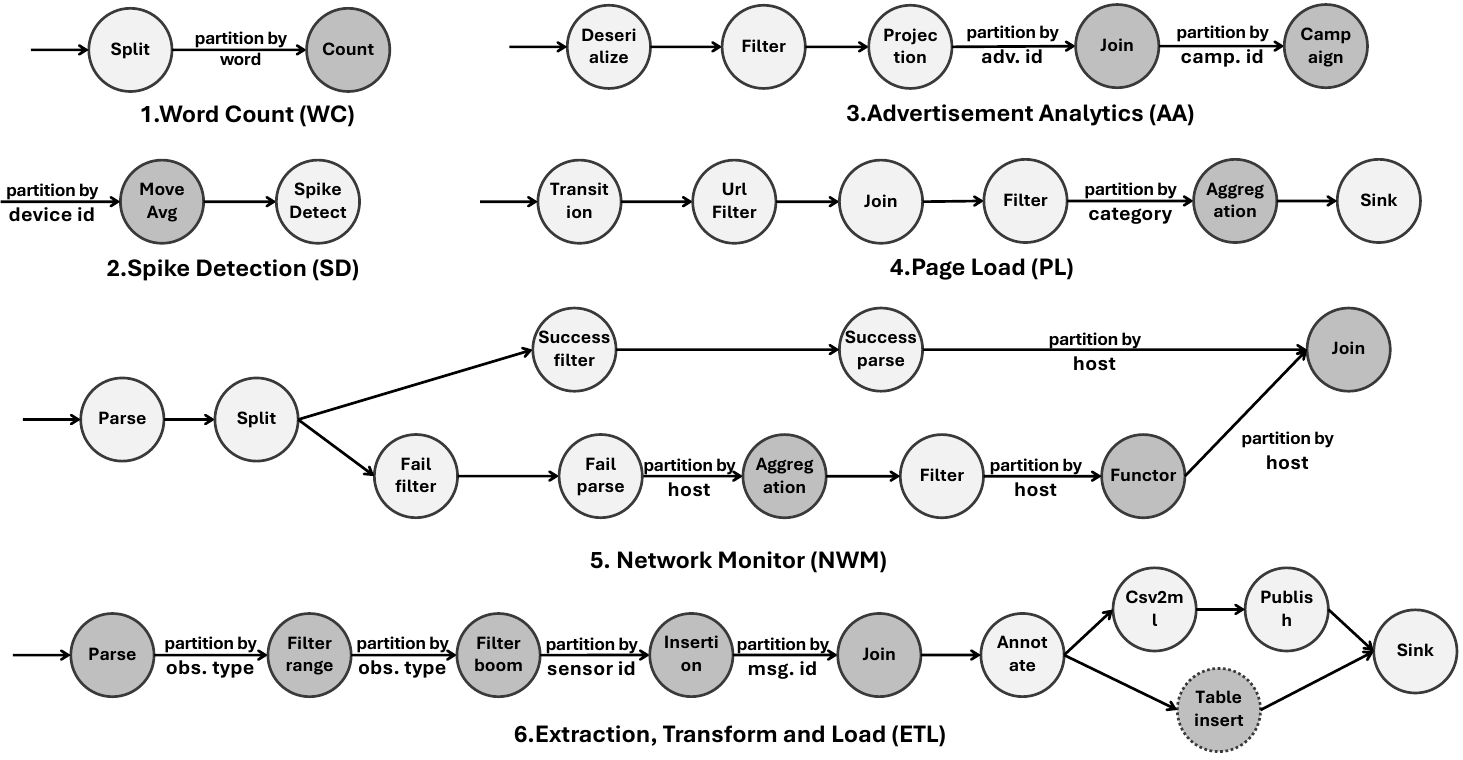}
\caption{Benchmark Application Topologies. Stateless operators are shown in light grey, partitioned stateful operators in dark grey, and stateful operators in dark grey with a dotted outline.}
\label{fig:3.0.application_topology}
\end{figure*}

Our primary goal is to increase throughput and reduce end-to-end latency. We define \emph{throughput} as the number of requests processed end-to-end (i.e., by all operators) per second, while \emph{end-to-end latency} refers to the time it takes for a request and all of its subsequent chained requests to be processed by the entire pipeline, encompassing both scheduling and execution time. Each experiment was run five times, with each run lasting for ten minutes. The results presented in this paper are the average of these five runs. As shown in Fig.~\ref{fig:3.0.application_topology}, we evaluate using the following applications: 

\noindent\textbf{Word Count (WC)} counts the occurrences of each unique word from text sources (2 operators)~\cite{bordin2020dspbench}.

\noindent\textbf{Spike Detection (SD)} maintains a moving window of temperature readings for each sensor ID and uses them to detect whether a newly received event is a spike (2 operators)~\cite{bordin2020dspbench}.

\noindent\textbf{Advertisement Analytics (AA)} maps advertisements to their corresponding campaigns and counts the number of events per campaign within a window (5 operators)~\cite{chintapalli2016yahoobenchmarking}.

\noindent\textbf{Page Load (PL)} computes the average page load time per page category by analyzing a stream of log files (6 operators)~\cite{xu2016stela}.

\noindent\textbf{Network Monitor (NWM)} monitors log files and looks for successful logins that were preceded by multiple failed attempts from the same host (10 operators)~\cite{gedik2014elastic}.

\noindent\textbf{Extraction, Transform and Load (ETL)} processes data streams through filtering, interpolation, and metadata annotation, and then persists the results (10 operators)~\cite{shukla2017riotbench}.

Apart from Advertisement Analytics (AA), the input workloads for all applications are characterized by irregular key distributions, leading to skewed loads across partitioned stateful operator instances.

\subsection{Request Transmission Time}

This experiment demonstrates a reduction in end-to-end latency achieved by enhancing request locality. To isolate end-to-end latency from the effects of batching and resource contention, this experiment sets the batch size to one and processes only a single input request at a time. In this experiment, our method is evaluated without runtime reconfiguration and compared with two baseline approaches based on median and 95th-percentile end-to-end latency.

\begin{figure}[]
\centering
\includegraphics[width=0.45\textwidth]{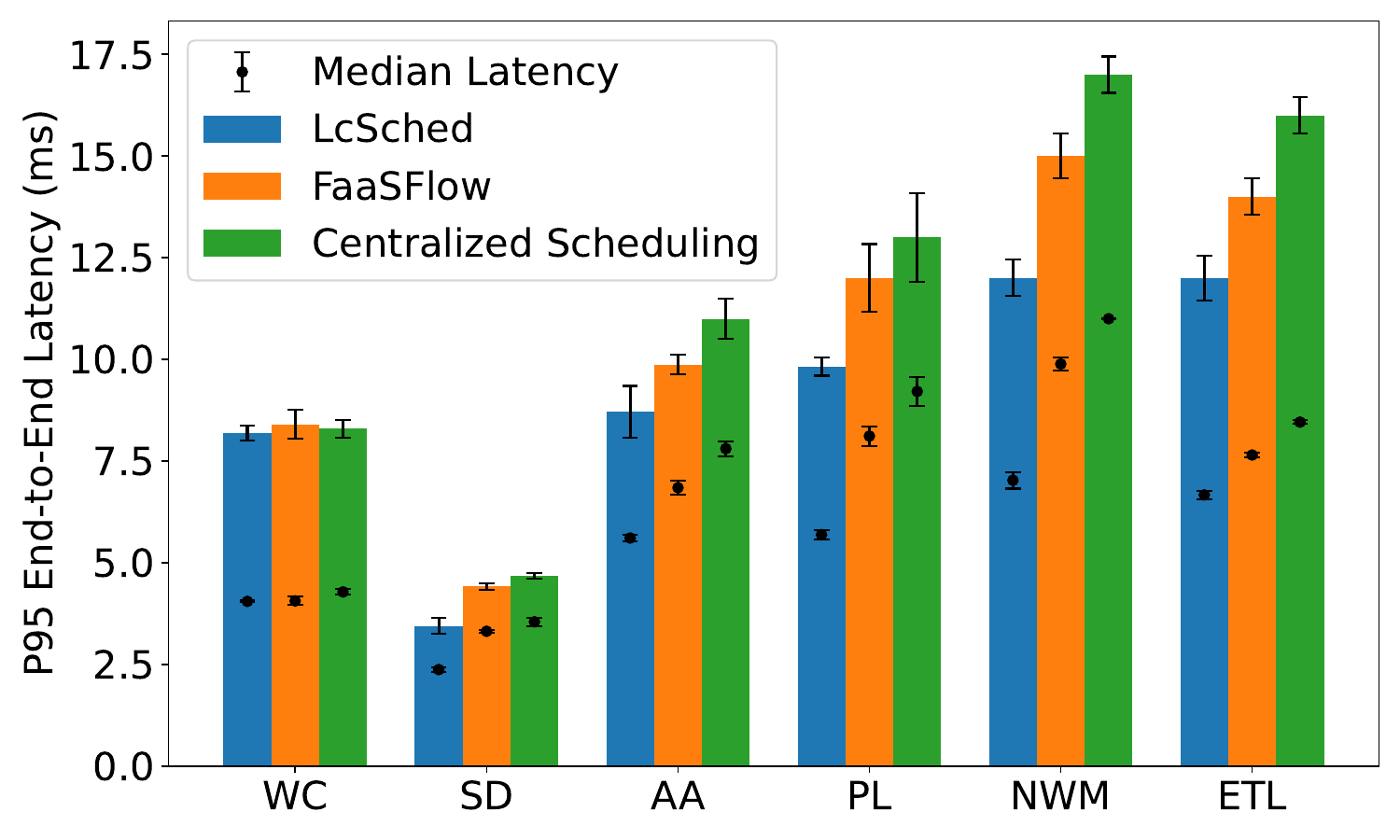}
\caption{Comparison of end-to-end latency (median and p95) across different scheduling methods.}
\label{fig:4.1.trans_comparision}
\end{figure}

As shown in Fig.~\ref{fig:4.1.trans_comparision}, our methods achieved lower end-to-end latency than FaaSFlow for most applications, and FaaSFlow, in turn, outperformed the centralized scheduling baseline. In WC, all three methods exhibit similar latencies. Our approach performs similarly to FaaSFlow due to the characteristics of WC: the first operator is stateless whose input attributes only contain sentence, the second operator is a partitioned stateful operator partitioned by word. As a result, our method splits these operators into two groups, providing no reduction in cross-worker request transfers compared to FaaSFlow. The centralized baseline's performance is also comparable. Since the first operator in WC generates approximately 10 chained requests, which are processed in parallel. The end-to-end latency, measured until the last of these requests completes, is therefore dominated by the variance in their parallel execution times, masking the relatively small additional overhead of the centralized scheduler.

The SD application contains two operators, and each input request to the first operator generates exactly one chained request. Our method executes every chained request locally, whereas FaaSFlow executes each one cross-worker. As a result, the end-to-end latency becomes a clearer indicator of variations in request transmission time. In SD, the median latency of our method, FaaSFlow and centralized scheduling is 2.4, 3.3, 3.7 ms, respectively. These results show that, for a single chained request, local execution is approximately 0.9 ms faster than cross-worker execution and 1.3 ms faster than routing through a centralized scheduler. 

We also observe that executing a higher proportion of chained requests locally leads to greater reductions in latency. The PL application, which consists of six operators and five data flows, serves as a clear example. For this application, our method executes the chained requests for four of the five data flows locally, whereas FaaSFlow routes all of them cross-worker. This difference in routing reduces the median end-to-end latency by 2.5 ms in favor of our approach.

\subsection{Scheduling Overhead - CPU usage}

In LcSched, each worker runs a scheduling engine alongside the executor pool, sharing the CPU resources. This experiment quantifies the CPU contention between the per-worker scheduling engine and executors. Measured in an unthrottled input stream, the reported results represent the mean utilization of each component, averaged across all workers.

\begin{table}[h]
  \centering
  \caption{CPU Utilization by Component (\%)}
  \label{tab:cpu_utilization_flat}
  \begin{tabular}{l cccccc}
    \toprule
    \textbf{Component} & \textbf{WC} & \textbf{SD} & \textbf{AA} & \textbf{PL} & \textbf{NWM} & \textbf{ETL} \\
    \midrule
    Executor    & 55.7 & 83.1 & 58.2 & 63.1 & 61.3  & 65.9 \\
    Scheduler    & 0.91  & 0.96  & 0.90  & 1.28  & 1.56  & 1.52  \\
    \bottomrule
  \end{tabular}
\end{table}

The result reflects the CPU time apportionment between executor computation and the scheduling of chained requests. For NWM, and ETL applications, each with ten operators, the scheduling engine's CPU consumption reaches around 2.5\% of the executor CPU consumption. Applications with fewer operators show a smaller ratio of scheduling to execution consumption. For example, the SD application exhibits a lower proportion of only 1.1\%, owing to its simpler structure with just two operators. Each input record from the source to SD is processed by just 2 executors and passed through 1 chained request. When the application is larger, such as the NWM application, it requires each input record to be processed by 10 executors and passed through 9 chained requests, resulting in a higher proportion of scheduling overhead.

Although the overall CPU utilization for the WC application appears similar to that of other applications, its scheduling overhead is highly concentrated. This is due to its fan-out structure: the first operator generates approximately 10 chained requests per input, while the second operator generates none. On the worker exclusively executing the first operator, we observed an executor-to-scheduler CPU utilization ratio of roughly 10:1. Under these conditions, the CPU usage of the scheduler must be taken into account.

\subsection{Runtime Reconfiguration}

\begin{figure*}[]
    \centering 
    % --- Top Row ---
    \begin{subfigure}[b]{0.32\textwidth}
        \centering
        \includegraphics[width=\textwidth]{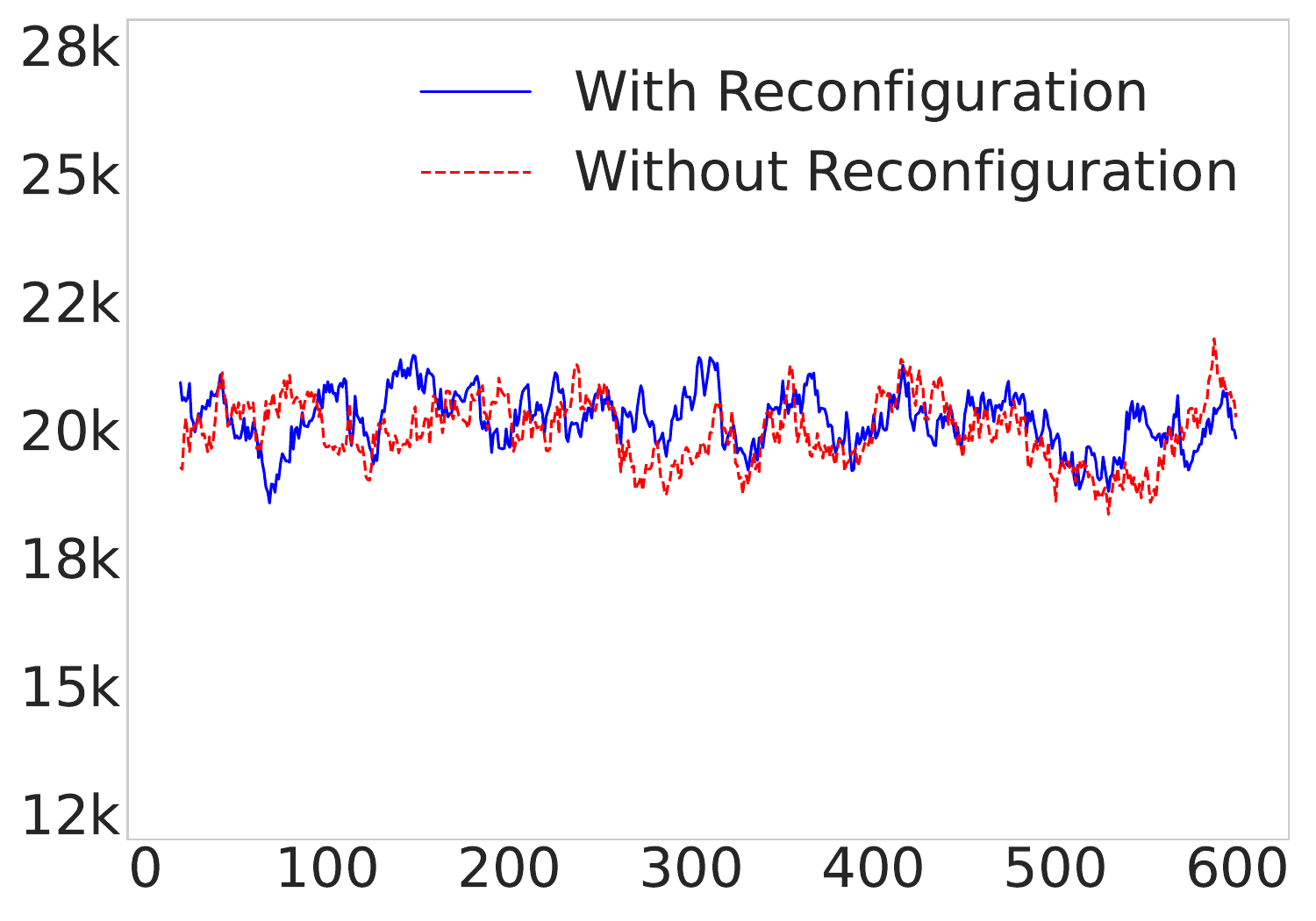}
        \caption{WC}
        \label{fig:4.2.wc}
    \end{subfigure}
    \hfill
    \begin{subfigure}[b]{0.32\textwidth}
        \centering
        \includegraphics[width=\textwidth]{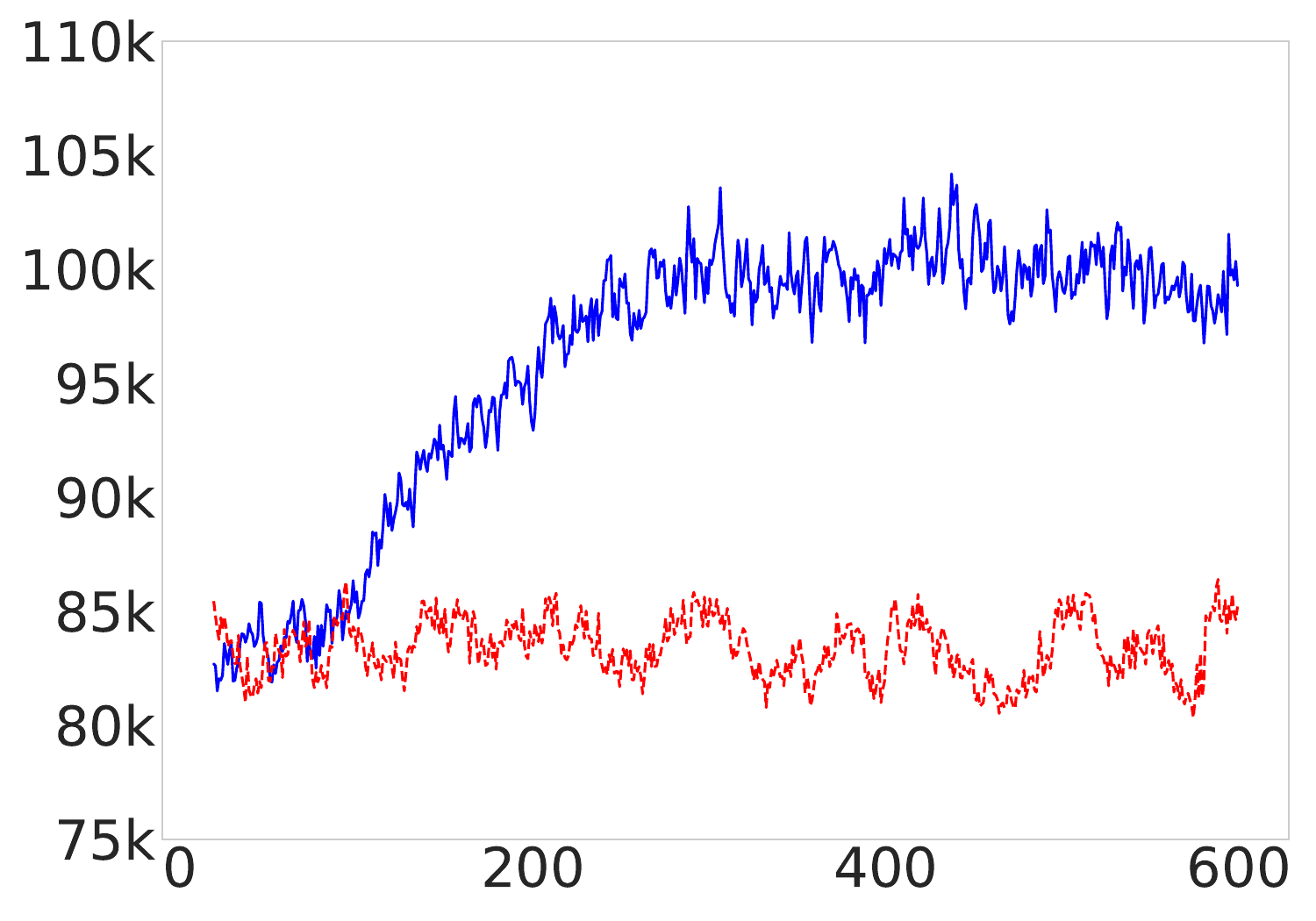}
        \caption{SD}
        \label{fig:4.2.sd}
    \end{subfigure}
    \hfill
    \begin{subfigure}[b]{0.32\textwidth}
        \centering
        \includegraphics[width=\textwidth]{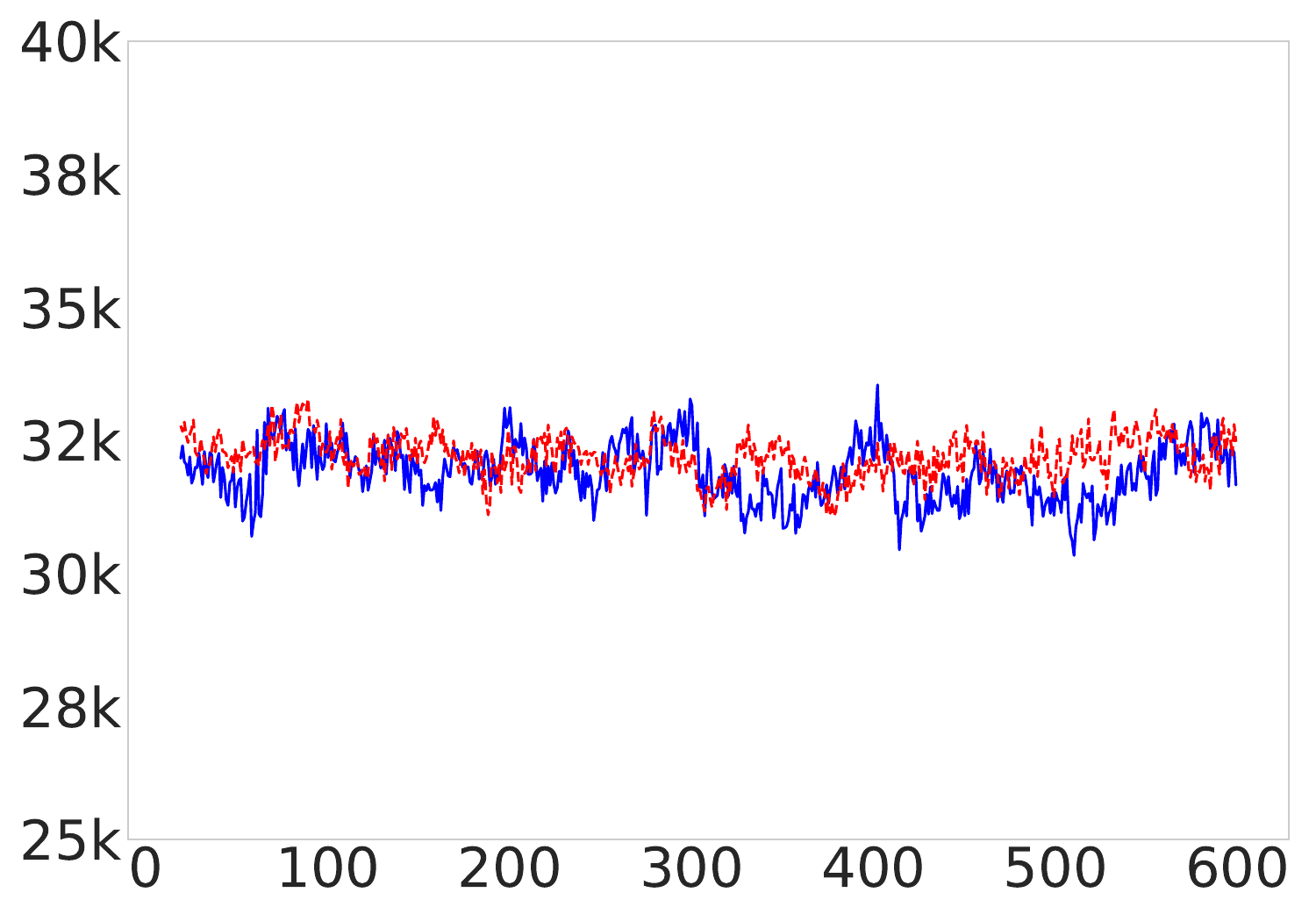}
        \caption{AA}
        \label{fig:4.2.aa}
    \end{subfigure}

    % --- Bottom Row ---
    \begin{subfigure}[b]{0.32\textwidth}
        \centering
        \includegraphics[width=\textwidth]{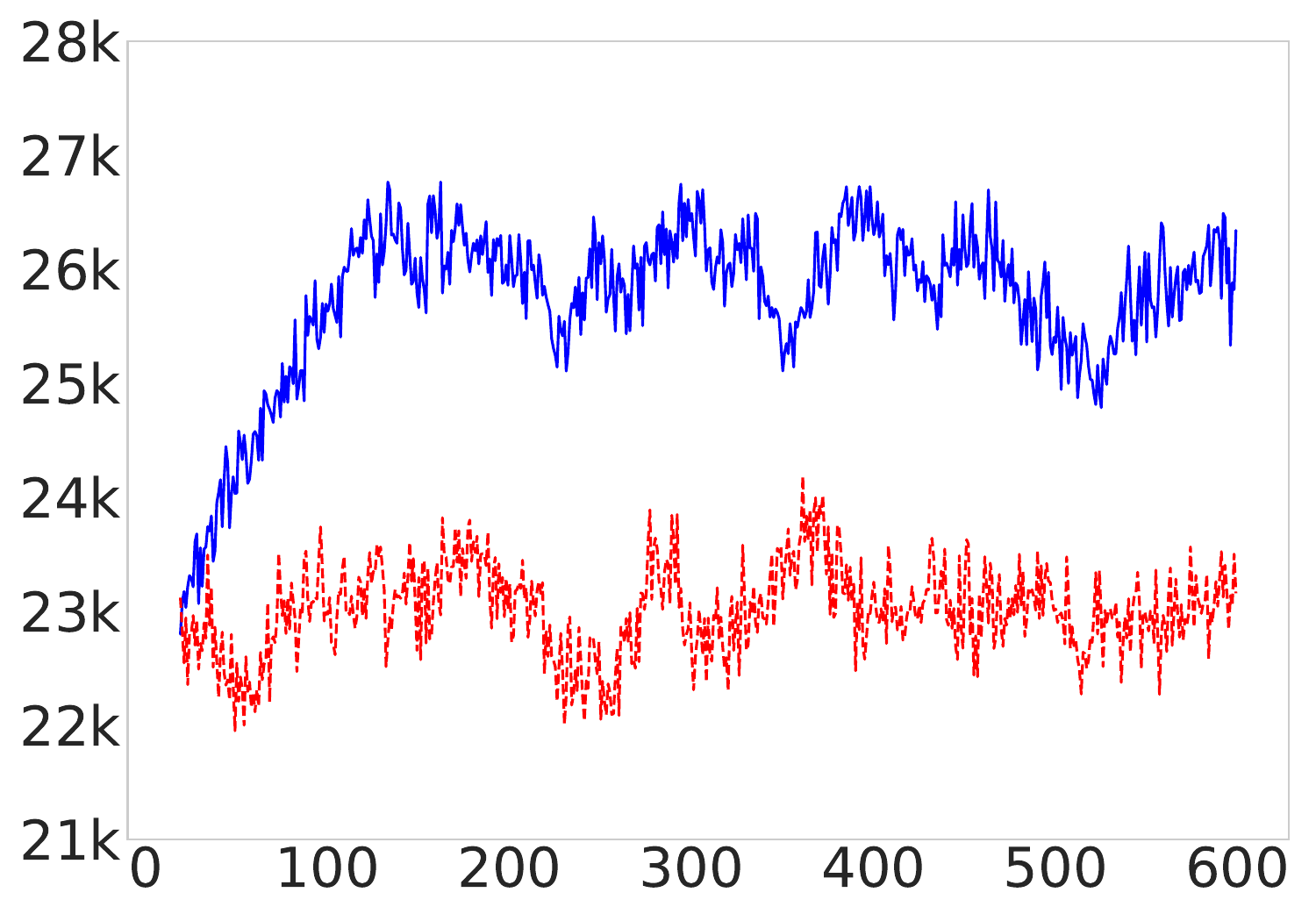}
        \caption{PL}
        \label{fig:4.2.pl}
    \end{subfigure}
    \hfill
    \begin{subfigure}[b]{0.32\textwidth}
        \centering
        \includegraphics[width=\textwidth]{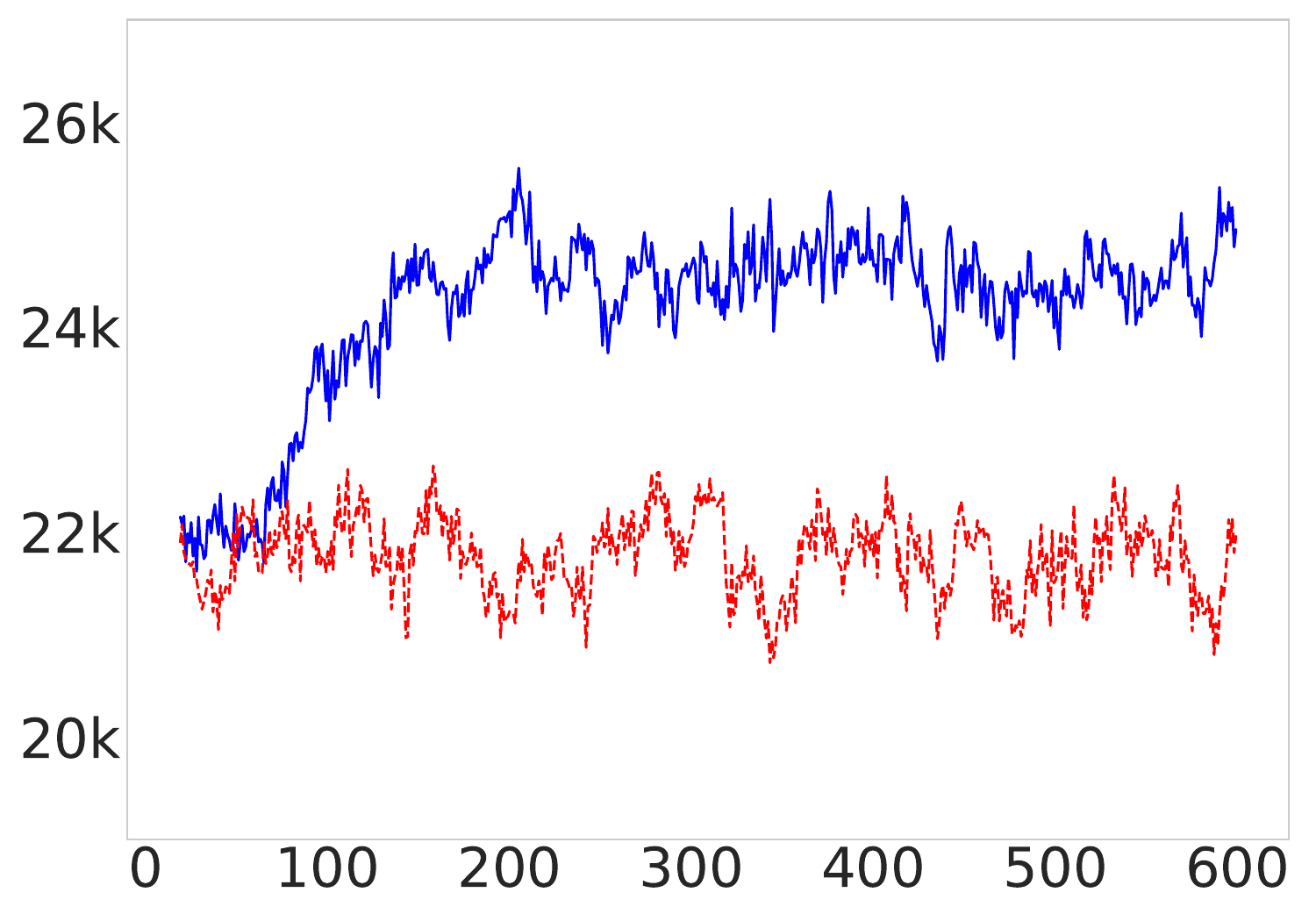}
        \caption{NWM}
        \label{fig:4.2.nwm}
    \end{subfigure}
    \hfill
    \begin{subfigure}[b]{0.32\textwidth}
        \centering
        \includegraphics[width=\textwidth]{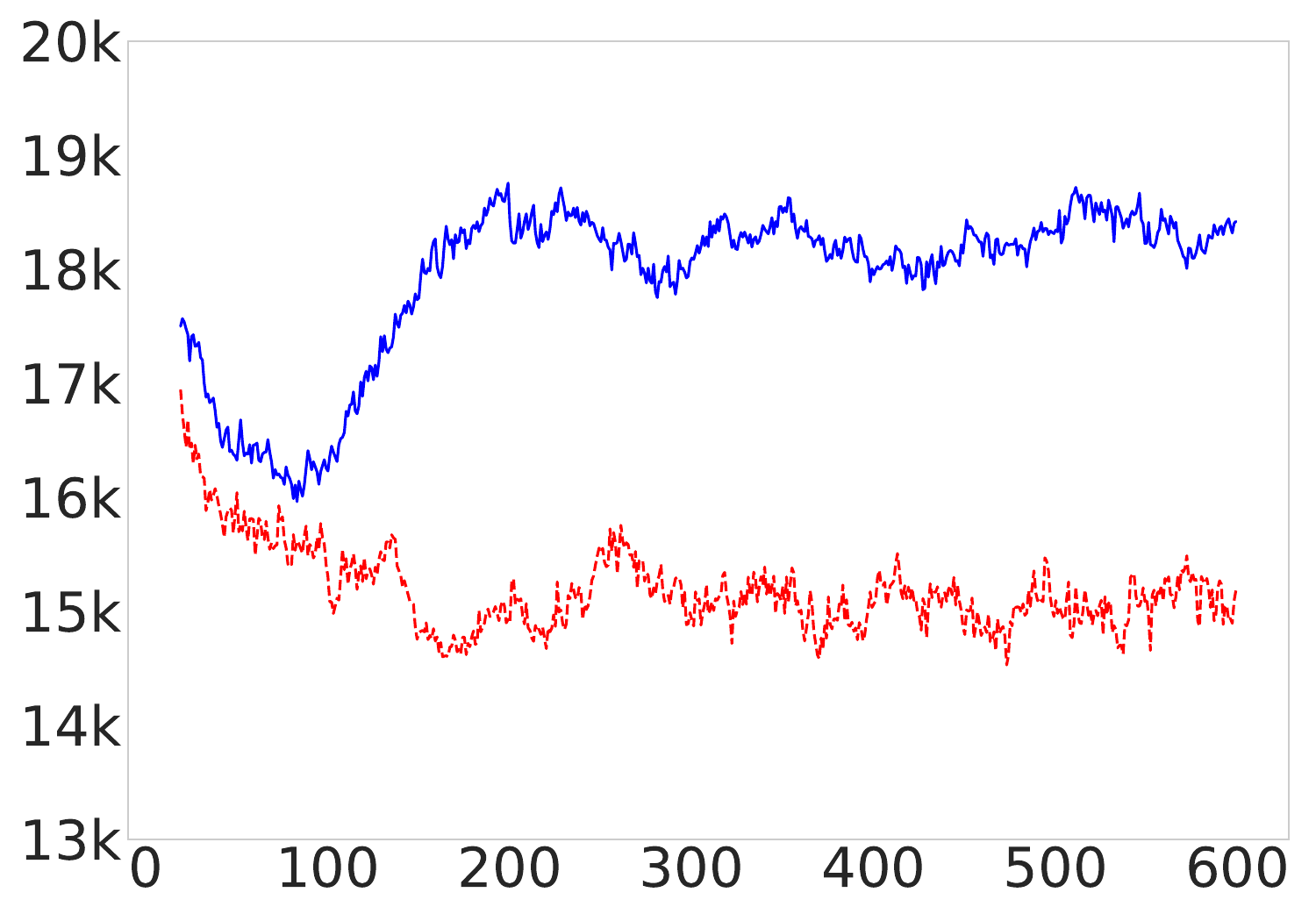}
        \caption{ETL}
        \label{fig:4.2.etl}
    \end{subfigure}

    \caption{Comparison of throughput over time with (blue line) and without (red line) runtime reconfiguration (smoothed with window size 20).}
    \label{fig:4.2.runtime_reconfig}
\end{figure*}

Our runtime reconfiguration mechanism balances the workload by strategically forwarding requests to underloaded workers. Without reconfiguration, requests are executed locally. This creates a trade-off between workload balance and request locality, which we explore in this experiment. We compare two configurations of our approach: one with reconfiguration enabled (reconfiguration period = 20s, $\eta$ = 0.05) and one without reconfiguration. By feeding the system an unthrottled input stream, we then record the throughput of both configurations over a 10-minute period.

As shown in Fig.~\ref{fig:4.2.runtime_reconfig}, for the SD, PL, NWM, and ETL applications, the two configurations initially exhibit similar throughput. However, as time passes, the dynamic reconfiguration begins to balance the workload. Consequently, the throughput of the reconfiguration-enabled configuration gradually increases and surpasses that of reconfiguration-disabled configuration. This result demonstrates that workload balancing can lead to higher overall throughput, even at the expense of sacrificing some request locality. For the ETL application, the throughput of both configurations exhibits a downward trend during the initial stage. This degradation occurs as the application’s state progressively expands, which increases the load on the state server and ultimately reduces throughput.

However, for the WC, AA application, runtime reconfiguration provides no performance improvement. The WC application contains no operators that are eligible for reconfiguration. In AA, all operators are already balanced without reconfiguration.

\subsection{Performance Comparison}

In this experiment, we compare the average throughput as well as the median and 95th-percentile latency achieved by each method. Our approach is configured with runtime reconfiguration, and all methods are evaluated under an unthrottled input stream to measure peak performance.

\begin{figure}[]
\centering
\includegraphics[width=0.45\textwidth]{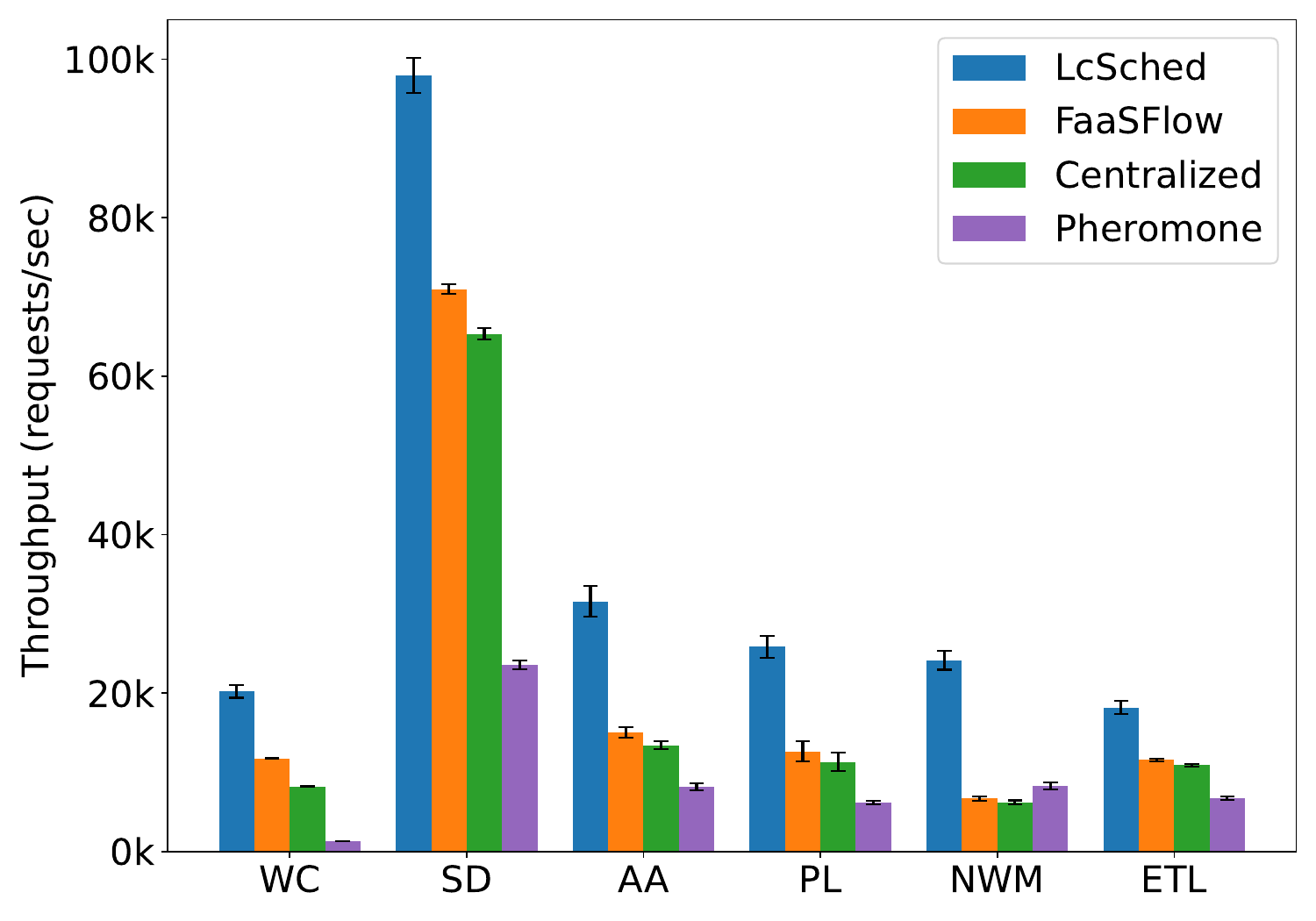}
\caption{Overall Throughput Comparison by Method.}
\label{fig:4.3.ten_worker_throughput}
\end{figure}

\begin{figure}[]
\centering
\includegraphics[width=0.45\textwidth]{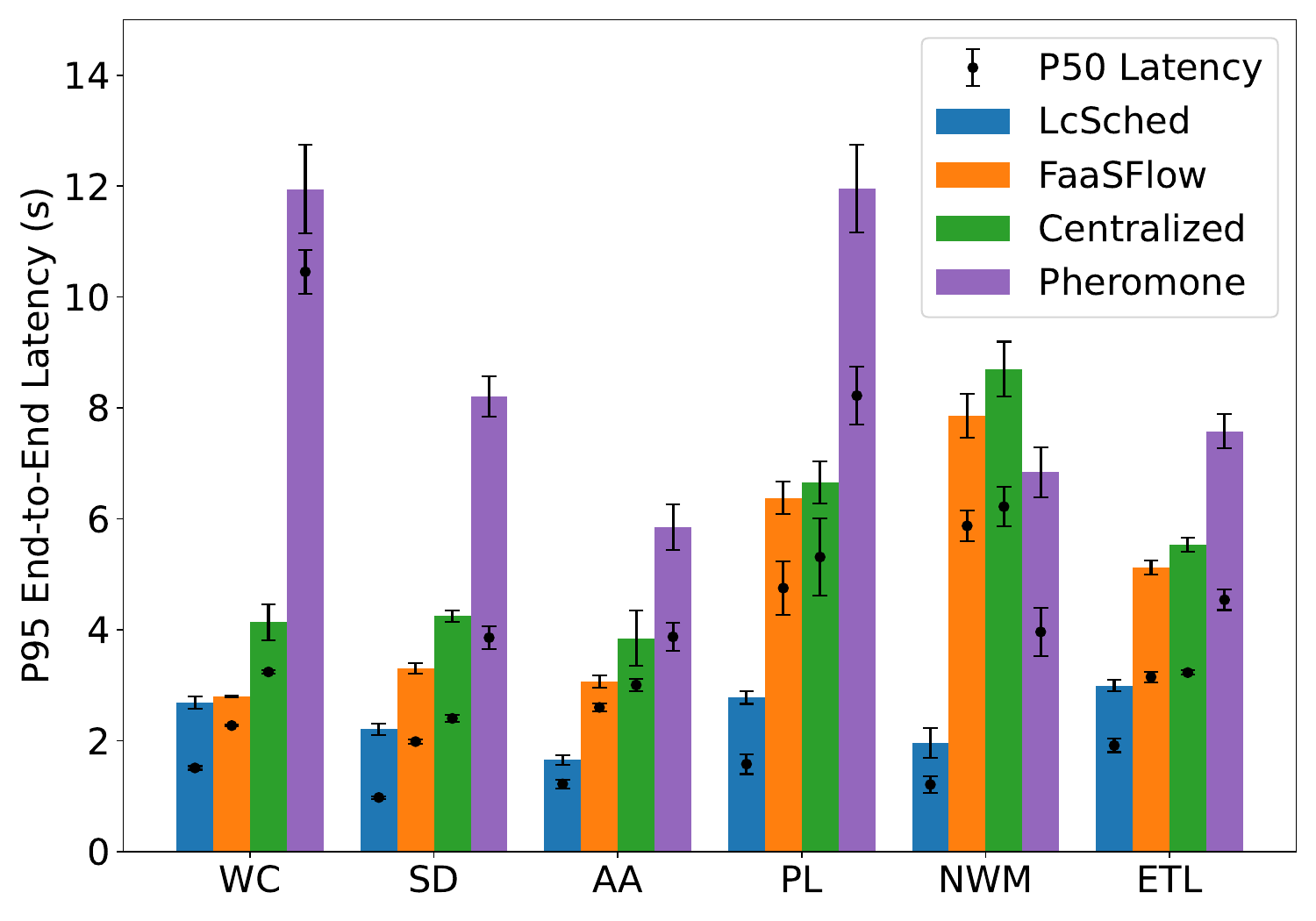}
\caption{Latency Comparison (median and p95) by Method.}
\label{fig:4.3.ten_worker_latency}
\end{figure}

As shown in Figs.~\ref{fig:4.3.ten_worker_throughput} and \ref{fig:4.3.ten_worker_latency}, our method consistently outperforms both baselines in all applications, achieving up to 3.6x higher throughput and up to 79\% lower median end-to-end latency than FaaSFlow, and up to 3.9x higher throughput and up to 81\% lower median latency than the centralized scheduling baseline. Compared with Pheromone, our method achieves up to 15.2x higher throughput and up to 85\% lower median latency. 

The throughput gains are correlated with the level of request locality that is achieved in our method. In the NWM application, where 90\% of chained requests are executed locally in LcSched, throughput improves by 3.6x compared to FaaSFlow. However, for the ETL application, where only 40\% of chained requests are executed locally in LcSched, the throughput improvement over FaaSFlow is a more modest 1.6x. For the WC application, our method does not improve request locality, yet it still delivers higher throughput. The improvement stems from allocating additional capacity to the first operator, which becomes a bottleneck in other methods due to the scheduling and transfer of numerous chained requests.

Pheromone performs worst as it relies on remote state access, which incurs high latency. Its performance is particularly poor for WC. While Pheromone tends to execute all chained requests on the worker where they originate, WC's first operator fans out each sentence into multiple words. Because sentence lengths differ widely in the input stream, this creates highly uneven bursts of word-count requests across workers, leading to severe workload imbalance. In contrast, Pheromone performs relatively better on NWM, as seven of the ten operators are stateless, reducing the impact of remote state accesses.

% Related work table
\begin{table*}[ht]
\centering
\caption{A comparison with related work.}
\label{tab:related_comparison}
\begin{tabular}{lllcccccc}
\hline
Representative Work & Architecture & Method & Environment & FC & LS & SL & RL & RA \\
\hline
Faasm~\cite{shillaker2020faasm}        & Centralized & Operator-based & Cloud & \ding{51} &           & \ding{51} &           &           \\
OFC~\cite{mvondo2021ofc}        & Centralized & Operator-based & Cloud & \ding{51} &           & \ding{51} &           & \ding{51} \\
Atoll~\cite{singhvi2021atoll}          & Semi-global & Operator-based & Cloud & \ding{51} &           & \ding{51} &           &  \ding{51} \\
Hermod~\cite{kaffes2022hermod}         & Semi-global & Operator-based & Cloud & \ding{51} &           &           &           & \ding{51} \\
Faastlane~\cite{kotni2021faastlane}    & Distributed & Stream-based  & Cloud & \ding{51} & \ding{51} &           & \ding{51} &           \\
Pheromone~\cite{yu2023pheromone}       & Distributed & Stream-based  & Cloud & \ding{51} & \ding{51} &           & \ding{51} &           \\
Cloudburst~\cite{sreekanti2020cloudburst} & Distributed & Operator-based & Cloud & \ding{51} &        & \ding{51} &           & \ding{51} \\
FaaSFlow~\cite{li2022faasflow}         & Distributed & Operator-based & Cloud & \ding{51} & \ding{51} & \ding{51} &           & \ding{51} \\
Ekko~\cite{chen2025ekko}               & Distributed & Operator-based & Edge  &           & \ding{51} & \ding{51} &           & \ding{51} \\
\hline
\textbf{Ours (LcSched)}                  & Distributed & Hybrid        & Cloud & \ding{51} & \ding{51} & \ding{51} & \ding{51} & \ding{51} \\
\hline
\end{tabular}
\\[1ex]
{\footnotesize FC: Fully-Connected Workers;\quad LS: Local Scheduling;\quad SL: State-Locality Aware;\quad RL: Request-Locality Aware;\quad RA: Runtime Adjustment}
\end{table*}

\subsection{Scalability}

To evaluate the scalability of different methods, we repeat the previous experiment on a 3-worker cluster. We keep all other parameters unchanged and compare the average throughput with the performance of the 10-worker cluster.

\begin{figure}[]
\centering
\includegraphics[width=0.45\textwidth]{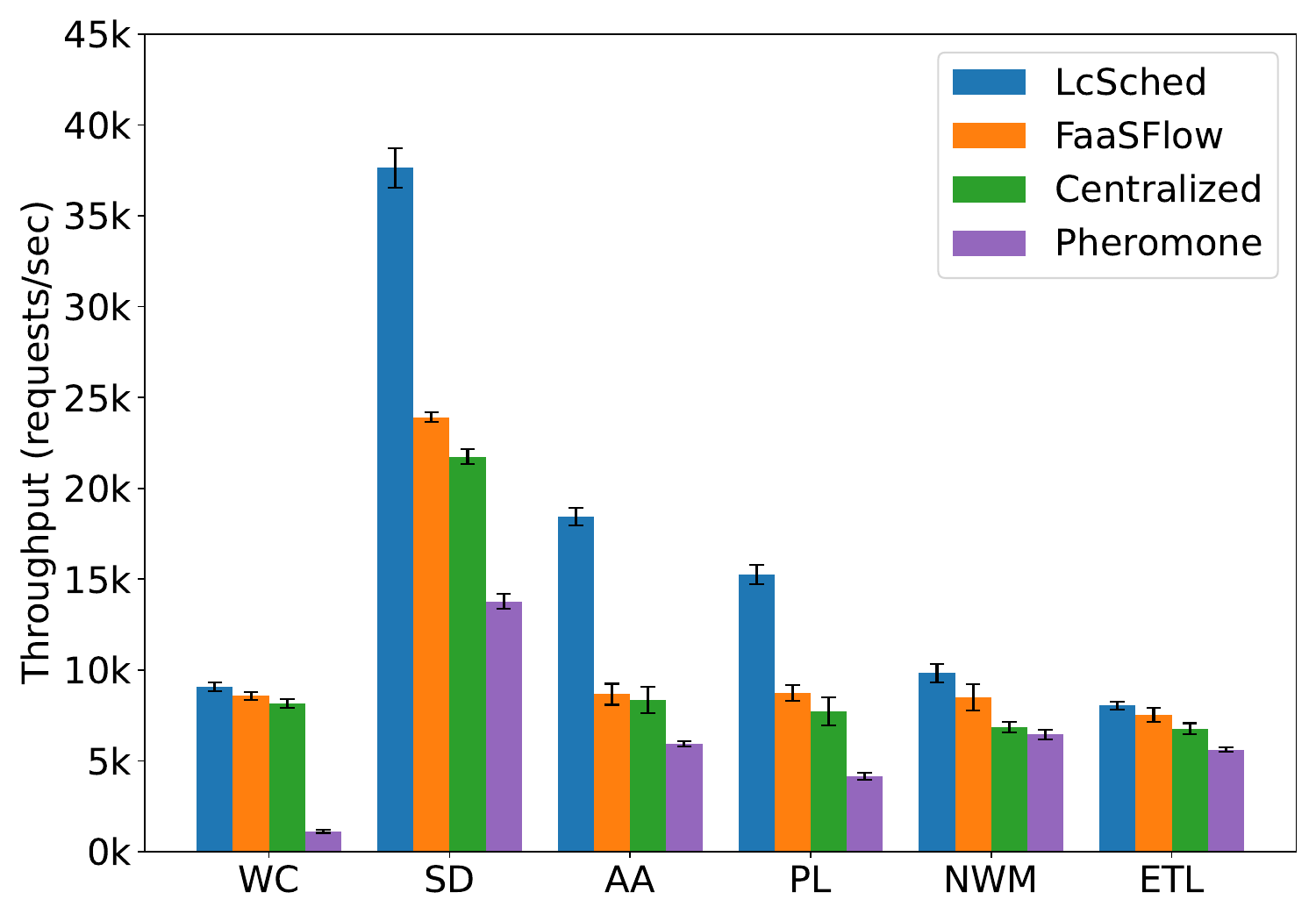}
\caption{Overall Throughput Comparison by Method (3 workers).}
\label{fig:4.4.three_worker_throughput}
\end{figure}

As shown in Fig.~\ref{fig:4.4.three_worker_throughput}, our method's performance advantage narrows on a smaller 3-worker cluster. Since FaaSFlow schedules more operators on the same worker in a small cluster, reducing cross-worker overhead, while the centralized baseline’s scheduler is less likely to become a bottleneck under the lighter traffic.

In WC, our method cannot increase request locality, which explains why it achieves similar throughput. When the cluster size increases to 10 workers, the throughput of the two baselines does not improve, whereas our method’s throughput rises from 9,000 to 20,000. Since these baseline methods do not account for worker load from scheduling and routing chained requests, they fail to allocate additional resources to the first operator as the cluster grows, creating a bottleneck.

Compared to FaaSFlow, our method achieves higher throughput for SD, AA, and PL applications, as it further improves the request locality. However, for larger applications such as NWM and ETL, Fig.~\ref{fig:2.2.faasflow_internode} shows that FaaSFlow routes most chained requests locally in the 3-worker cluster, explaining why the throughput of our method is similar in these cases. For the ETL application, scaling the cluster from 3 to 10 workers increases the throughput of our method from 8,100 to 18,100, while FaaSFlow’s throughput grows only from 7,500 to 8,500 due to increased cross-worker request transfers. The NWM application exhibits a similar pattern. These results demonstrate that the performance benefits of our method become more pronounced with larger cluster sizes and are applicable to applications with various operator scales. Pheromone consistently performs worst on the 3-worker cluster as it requires frequent remote state accesses.

\section{Related Work}
\label{sec:related_work}
A summary of related works and comparison of their features with our work is shown in Tab.~\ref{tab:related_comparison}.

\textbf{Stateful Serverless.} Existing stateful serverless systems can be broadly categorized into two types. The first type maintains a local state server on each worker node, allowing executors to access state locally and avoid cross-worker communication~\cite{shillaker2020faasm, mvondo2021ofc,li2022faasflow, lee2023flowkv}. The second type tailors remote storage servers for serverless~\cite{khandelwal2022jiffy, barcelona2023glider}. For instance, Glider~\cite{barcelona2023glider} integrates stateful near-data computation within the storage tier to reduce data movement between compute and storage tiers. However, inter-node communication remains inevitable in this type. Since stream processing applications invoke short-lived functions at high frequency with frequent state accesses, remote state access becomes a significant bottleneck. Therefore, LcSched builds upon the first type.

\textbf{Semi-global scheduler.} To alleviate the workload of centralized schedulers in large clusters, semi-global scheduling strategy has been proposed. This approach partitions the cluster into sub-clusters, each managed by a semi-global scheduler~\cite{saurez2021geodistributedscheduler, singhvi2021atoll, wang2021faasnet, kaffes2022hermod, patterson2022hivemind}. It can reduce the workload of the centralized scheduler, but all chained requests still need to be forwarded to another node for scheduling.

\textbf{Stream-based Scheduler.} To minimize the communication overhead of transferring chained requests between workers, data-aware distributed scheduling has been widely explored. These schedulers aim to execute chained requests on the same worker where they originate, or even within the same executor~\cite{akkus2018sand, kotni2021faastlane, jia2021nightcore, wang2024briskchain}. Pheromone~\cite{yu2023pheromone} leverages shared memory to efficiently store and transfer intermediate data across chained functions, while preferentially executing chained requests locally to reduce data exchange costs. However, these methods do not account for the persistent operator state accessed during execution, which is distinct from the intermediate data carried between requests.

\textbf{Edge Environments.} Distributed scheduling has also been explored in edge computing, where devices run independent schedulers that route requests to neighbors based on resource availability~\cite{cicconetti2020decentralized, tang2022networkiotscheduling, chen2025ekko}. However, edge environments typically feature only partial connectivity between nodes, and operators are naturally constrained to specific nodes. In cloud environments, by contrast, full connectivity among workers without such placement constraints, enlarging the scheduling search space and making placement decisions more complex.

\textbf{Operator-based scheduler.} Some distributed scheduling studies follow an operator-based approach where operators are preassigned to specific nodes before the application starts~\cite{sreekanti2020cloudburst, li2022faasflow}. During execution, the distributed scheduler routes chained requests directly to the preassigned nodes. Pinning a stateful operator or a partitioned stateful operator to a specific worker inherently deploys its associated state on that worker. As a result, all requests for that operator are naturally collocated with their required state. However, pinning an operator to a specific worker undermines the flexibility of the serverless model, as it may force subsequent chained requests to be routed across workers for execution.

\section{Conclusions and Future Work}
\label{sec:conclude}
We proposed LcSched, a request and state locality-aware distributed scheduling strategy for stream processing applications in stateful serverless frameworks. LcSched achieves locality by partitioning applications into operator groups, which ensures that chained requests to operators within the same group can be executed and access state locally. Furthermore, LcSched performs dynamic load balancing by monitoring the runtime workload of each worker. Experimental results show the enhancements in both throughput and latency over the state-of-the-art distributed scheduling strategy. In the future work, we plan to adjust the number of workers dynamically to adjust the fluctuation of input workload.

\textbf{Source Code:} Our LcSched software can be accessed from Gitbub page: https://github.com/LcSched

\textbf{Declaration of Generative AI and AI-assisted technologies in the writing process:} During the preparation of this work, the first author used Claude to improve language and readability. After using the tools, the first author reviewed and edited the content as needed and takes full responsibility for the content of the publication.

% \textbf{CRediT authorship contribution statement
% }

% \textbf{Tianyu Qi}: Conceptualization, Writing – original draft, Software, Validation, Visualization, Software, Formal analysis.  
% \textbf{Maria A. Rodriguez} and \textbf{Rajkumar Buyya}: Conceptualization, Methodology, Writing – review \& editing, Supervision.

% \textbf{Declaration of competing interest}

% The authors declare that they have no known competing financial interests or personal relationships that could have appeared to
% influence the work reported in this paper.

\bibliographystyle{cas-model2-names}
\bibliography{reference_papers}

\vspace{12pt}
\end{document}